\documentclass[11pt,a4paper]{article}
\pdfoutput=1
\usepackage{jheppub}
\usepackage{graphicx,psfrag}
\usepackage{bm}
\usepackage{mathbbol,verbatim}
\usepackage{slashed}
\usepackage{graphics}
\usepackage{ulem}
\allowdisplaybreaks

\usepackage{tikz}
\usetikzlibrary{trees}
\usetikzlibrary{decorations.pathmorphing}
\usetikzlibrary{decorations.markings}
\usetikzlibrary{decorations, decorations.markings, decorations.pathmorphing, arrows, graphs, shapes.geometric, snakes}
\usetikzlibrary{arrows}

\tikzset{
photon/.style={decorate, decoration={snake}, draw=red},
dark/.style={draw=gray, postaction={decorate},
        decoration={markings,mark=at position .55 with {\arrow[draw=gray]{>}}}},
antidark/.style={draw=gray, postaction={decorate},
        decoration={markings,mark=at position .55 with {\arrow[draw=gray]{<}}}},
electron/.style={draw=violet, postaction={decorate},
        decoration={markings,mark=at position .55 with {\arrow[draw=violet]{>}}}},
positron/.style={draw=violet, postaction={decorate},
        decoration={markings,mark=at position .55 with {\arrow[draw=violet]{<}}}},
neutrino/.style={draw,color=violet,thick, postaction={decorate} },
neutrinolight/.style={draw=blue, postaction={decorate} },
quark/.style={draw=blue, postaction={decorate},
        decoration={markings,mark=at position .55 with {\arrow[draw=blue]{>}}}},
antiquark/.style={draw=blue, postaction={decorate},
        decoration={markings,mark=at position .55 with {\arrow[draw=blue]{<}}}},
heavyquark/.style={draw=purple, postaction={decorate},
        decoration={markings,mark=at position .55 with {\arrow[draw=purple]{>}}}},
antiheavyquark/.style={draw=purple, postaction={decorate},
        decoration={markings,mark=at position .55 with {\arrow[draw=purple]{<}}}},
        gluon/.style={decorate, draw=or,
        decoration={coil,amplitude=2pt, segment length=3pt}},
gluon/.style={decorate, draw=or,
        decoration={coil,amplitude=2pt, segment length=3pt}},
ZZ/.style={decorate, decoration={snake,amplitude=1.5pt, segment length=5pt}, draw=green},
left,
  }

\definecolor{green}{rgb}{0.03,0.84,0.13}
\definecolor{test}{rgb}{0.03,0.74,0.33}
\definecolor{viol}{rgb}{0.44,0,0.94}
\definecolor{or}{rgb}{0.95,0.65,0}

\begin{document}

\preprint{UMD-PP-016-008, ULB-TH/16-14}

\title{Naturally Stable Right-Handed Neutrino Dark Matter}

\author[a]{P. S. Bhupal Dev,}
\author[b]{Rabindra N. Mohapatra,}
\author[c,d]{Yongchao Zhang}
\affiliation[a]{ Max-Planck-Institut f\"{u}r Kernphysik, Saupfercheckweg 1, D-69117 Heidelberg, Germany}
\affiliation[b]{Maryland Center for Fundamental Physics, Department of Physics, University of Maryland, College Park, MD 20742, USA}
\affiliation[c]{Service de Physique Th\'{e}orique, Universit\'{e} Libre de Bruxelles, Boulevard du Triomphe, CP225, 1050 Brussels, Belgium}
\affiliation[d]{School of Physics, Sun Yat-Sen University, Guangzhou 510275, China}

\emailAdd{bhupal.dev@mpi-hd.mpg.de}
\emailAdd{rmohapat@umd.edu}
\emailAdd{yongchao.zhang@ulb.ac.be}


\abstract{
  We point out that a class of non-supersymmetric models  based on the gauge group $SU(3)_C \times SU(2)_L\times SU(2)_R\times U(1)_{Y_L}\times U(1)_{Y_R}$ possesses an automatic, exact $Z_{2 }$ symmetry under which the fermions in the $SU(2)_R\times U(1)_{Y_R}$ sector (called $R$-sector) are odd and those in the  $SU(2)_L\times U(1)_{Y_L}$  sector (called $L$-sector or the Standard Model sector) are even. This symmetry, which is different from the usual parity symmetry of the left-right symmetric models, persists in the lepton sector even after the gauge symmetry breaks down to $SU(3)_C \times U(1)_{\rm EM}$. This keeps the lightest right-handed neutrino naturally stable, thereby allowing it to play the role of dark matter (DM) in the Universe. There are several differences between the usual left-right models and the model presented here: (i) our model can have two versions, one which has no parity symmetry so that the couplings and masses in the $L$ and $R$ sectors are unrelated, and another which has parity symmetry so that couplings are related but with an extra Higgs doublet in each sector so that the fermion mass patterns are different; (ii) the $R$-sector fermions are chosen much heavier than the $L$-sector ones in both scenarios; and finally (iii) both light and heavy neutrinos are Majorana fermions with  the light neutrino masses arising from a pure type-II seesaw mechanism. We discuss the DM relic density, direct and indirect detection prospects and associated collider signatures of the model. Comparing with current collider and direct detection constraints, we find a lower bound on the DM mass of order of 1 TeV.  We also point out a way to relax the DM unitarity bound in our model for much larger DM masses by an entropy dilution mechanism. An additional feature of the model is that the DM can be made very long lived, if desired,  by allowing for weak breaking of the above $Z_{2}$ symmetry. Our model also predicts the existence of long-lived colored particles which could be searched for at the LHC. }


\maketitle
\section{Introduction}
The existence of dark matter (DM) constituting about 26\% of the energy budget of the Universe is by now well established from astrophysical and cosmological observations~\cite{Ade:2015xua}. It is also well known that understanding it requires the existence of new particle(s) beyond the Standard Model (SM) with very specific properties~\cite{Bertone:2004pz}. For instance, since the DM must be either absolutely stable or very long lived (i.e. lifetime longer than the age of the Universe), it implies that there must be an exact symmetry (or a very weakly broken symmetry) in the extended model under which all the SM particles are even while the DM particle is odd.\footnote{The most widely discussed example of such a discrete symmetry is the $R$-parity in supersymmetric models~\cite{Farrar:1978xj}.} This symmetry not only makes the DM stable but also allows it to annihilate only in pairs to create its observed relic density in the Universe. If evidence for the DM decaying emerges in future data, one can accomplish this using a soft breaking of this symmetry.  Another property of the DM which is necessary for it not to over-annihilate in the early Universe is that it must be neutral under  $SU(3)_C$ color and $U(1)_{\rm EM}$ electric charge. This color and charge neutrality property of DM naturally brings to mind one of the beyond SM candidates that is widely discussed in connection with neutrino physics, i.e. the right-handed neutrino (RHN), henceforth denoted by $N$. In a bottom-up approach, its mass can in principle be anywhere from eV to $10^{14}$ GeV or so. However, to constitute 100\% of the DM its mass should be above 0.4 keV just from the fact that being a fermion, its phase space distribution in a galaxy cannot exceed that of the degenerate Fermi gas~\cite{Tremaine:1979we, Boyarsky:2008ju, Gorbunov:2008ka}. The lower bound becomes slightly stronger $\gtrsim 1$ keV after taking into account a particular production mechanism and the corresponding primordial distribution of RHNs~\cite{Dodelson:1993je, Shi:1998km, Abazajian:2001nj}. The important thing to note here is that the keV-scale RHN can be made cosmologically stable without the need for any discrete symmetry, since all SM particles, except the photon and active neutrinos, are heavier and the lifetime of the radiative decay $N\to \gamma \nu$ will be sufficiently long for an appropriately small active-sterile neutrino mixing parameter, while being consistent with the neutrino oscillation data. A concrete realization of this scenario is the $\nu$MSM~\cite{Asaka:2005an, Canetti:2012kh}. The keV-scale RHN as the DM and its astrophysical implications have  been widely studied~\cite{Abazajian:2001vt, Abazajian:2012ys, Adhikari:2016bei}, but unfortunately this scenario is testable neither in the conventional DM direct detection nor in collider experiments, but only via its X-ray line signal from $N\to \gamma \nu$.\footnote{We note in passing that the unidentified line at 3.5 keV from XMM-Newton and Chandra X-ray observations~\cite{Bulbul:2014sua, Boyarsky:2014jta}, widely attributed to a possible sterile neutrino DM signal~\cite{Adhikari:2016bei}, has {\it not} been confirmed by the latest Hitomi data~\cite{Aharonian:2016gzq}.} In this paper, we would like to address the question whether a heavy RHN with mass in the TeV range, accessible at current and future laboratory experiments, can constitute the DM content of the Universe and can arise in a particle physics model of neutrino mass in natural manner.

A popular class of models for heavy RHNs that explain neutrino mass are based on the type-I seesaw mechanism~\cite{type1a, type1b, type1c, type1d, type1e} where the RHNs couple to the SM lepton doublet via the SM Higgs doublet through the Yukawa interaction. The minimal left-right (LR) models~\cite{LR1, LR2, LR3} provide a natural example of ultraviolet (UV) completion for type-I seesaw~\cite{type1b} where the RHNs are required to exist by consistency of the theory. In this kind of models, if the RHN  masses are in the 100 GeV to multi-TeV (or even higher mass) range, they will have rapid decays to SM leptons and the Higgs boson mediated by Yukawa couplings, or to SM fermions mediated by purely gauge interactions, and therefore, cannot be a viable DM candidate. There are, however, two exceptions to this statement:\footnote{A third possibility is to introduce a  new  fermion  (scalar)
multiplet with even (odd) $B-L$ charge, so that its lightest component is stabilized by the remnant $Z_2$ symmetry from $B-L$ breaking~\cite{Heeck:2015qra, Garcia-Cely:2015quu}.} (i) One can always choose to decouple one of the RHNs from the seesaw formula and make it stable or long lived such that it becomes a DM candidate~\cite{dibari, Fiorentin:2016avj}; however in this case, one has to choose the corresponding Yukawa coupling to be very small $\lesssim 3\times 10^{-26} ({\rm GeV}/{M_N})^{1/2}$ and assume that the RHN has no gauge interactions.
Alternatively, if the RHN has gauge interactions, they are only with a  $U(1)_{B-L}$ gauge boson so that its couplings to SM fermions are gauge-diagonal and an additional $Z_2$ symmetry is imposed to forbid the Yukawa couplings and make it stable; see e.g.~Refs.~\cite{Okada:2010wd, Kanemura:2011vm, Okada:2012sg, Lindner:2013awa, Basak:2013cga, Sanchez-Vega:2015qva, Biswas:2016bfo, Duerr:2015wfa, Okada:2016gsh, Ho:2016aye, Kaneta:2016vkq} for such RHN DM models based on the $SU(2)_L\times U(1)_Y\times U(1)_{B-L}$ gauge group. (ii) A second possibility in seesaw models is to have one of the RHNs decouple from the seesaw formula and remain light (with keV mass) as a result of a discrete flavor symmetry, e.g. $L_e-L_\mu-L_\tau$~\cite{rnm, Shaposhnikov:2006nn, merle}, $A_4$~\cite{Barry:2011wb, Barry:2011fp} or $Q_6$~\cite{Araki:2011zg}, and thus it becomes a warm DM~\cite{Adhikari:2016bei} in the Universe. One could of course have a RHN in a type-I seesaw framework with keV mass as in the case of $\nu$MSM~\cite{Asaka:2005an, Canetti:2012kh} and have it as a DM as noted above.

A key model building issue in all DM models is whether there is a symmetry that keeps the DM naturally stable, and if this symmetry needs to be imposed by hand (like $R$-parity in case of MSSM) or is automatic by the gauge structure and matter content of the model. For example, in the supersymmetric case, extension of the MSSM to include $U(1)_{B-L}$ gauge symmetry can make $R$-parity an automatic symmetry~\cite{RP1, RP2, RP3}. In this paper, we discuss a class of non-supersymmetric gauge models based on the gauge group $SU(3)_C \times SU(2)_L\times SU(2)_R\times U(1)_{Y_L}\times U(1)_{Y_R}$  where there appears an automatic $Z_2$ symmetry that keeps the lightest RHN stable, thereby making it a natural DM candidate. Similar models for RHN DM have been discussed before in terms of its subgroup $SU(3)_C \times SU(2)_L \times SU(2)_R\times U(1)_{\cal Y}$~\cite{bem,dkmtz}. However in these models, the DM is unstable and its decay to SM leptons was in fact used to explain the  anomalous positron excess at GeV scale~\cite{bem} or the PeV neutrino excess at IceCube~\cite{dkmtz}. The $Z_2$ symmetry appears in these models only when some parameters allowed by the gauge symmetry are set to zero, whereas in the model presented here, the $Z_2$ symmetry is present for all values of parameters allowed by the extended gauge symmetry of the model. In that sense, the $Z_2$ symmetry that stabilizes the DM in our model is an automatic symmetry. Another difference in the model presented here from those of Refs.~\cite{bem,dkmtz} is that we keep the DM mass in the multi-TeV range, which has distinct implications for cosmology, direct and indirect detection, as well as collider searches. In fact, we note that combining collider and direct detection bounds, we can put a lower bound on the DM mass of order of 1 TeV. There is also an upper limit from unitarity arguments, which ranges from multi-TeV to PeV in our model, depending on the dilution mechanism. 
Another important feature of this class of models is that they exclusively lead to a type-II seesaw~\cite{type2a, type2b, type2c, type2d}  for neutrino masses, unlike in the minimal LR models, where both type-I and type-II seesaw mechanisms are inherently present and one needs to do some fine-tuning to switch off either of them. 

This paper is organized as follows: in Section~\ref{sec:model}, we present the details of the model. In Section~\ref{sec:density} we calculate the relic density of RHN DM. In Section~\ref{sec:4}, we discuss the direct detection constraints, and in Section~\ref{sec:5} the indirect detection constraints. In Section~\ref{sec:6}, we point out some LHC signals of the model. In Section~\ref{sec:7}, we briefly discuss other implications of our model, summarize our results and conclude. In Appendix~\ref{app:A}, we give the analytic expressions for calculating the annihilation cross sections for the relic density.

\section{Model}\label{sec:model}
In this section, we present two versions of the model, one where parity is a good symmetry relating the Yukawa couplings of the light and heavy sectors, and another where parity symmetry is explicitly absent from the beginning. In our subsequent discussion of phenomenology, we will focus only on the model without parity. 
\subsection{Model without parity symmetry}\label{sec:2.1}
The model is based on the gauge group $SU(3)_C \times SU(2)_L \times SU(2)_R \times U(1)_{Y_L} \times U(1)_{Y_R}$ without exact parity under which $L \leftrightarrow R$. In addition to the SM fermions which transform under the $SU(2)_L \times U(1)_{Y_L}$ symmetry,
we have a heavy analog which transforms nontrivially under the symmetry $SU(2)_R \times U(1)_{Y_R}$. In particular, the light fermions consist of the SM doublets $Q_L$, $\psi_L$ and the singlets $u_R$, $d_R$, $e_R$, which are collected in the upper left column of Table~\ref{tab:fields}, with the symmetry $SU(2)_L \times U(1)_{Y_L}$ being clearly the standard electroweak (EW) gauge symmetry. We will call this the $L$ (light or left-handed) sector of the model. In the $R$ (right-handed) sector, we have the heavy analog of the SM fermions: the $SU(2)_R$ doublets $\mathcal{Q}_{R}$, $\Psi_{R}$ and the singlets $\mathcal{U}_{L}$, $\mathcal{D}_{L}$, $\mathcal{E}_{L}$, which are collected in the upper right column of Table~\ref{tab:fields}. It is clear that both the SM and heavy fermionic sectors share the common $SU(3)_C$  symmetry, and the electric charge formula is given by 
\begin{eqnarray}
Q \ = \ I_{3L}+I_{3R}+\frac{1}{2} (Y_L+Y_R) \,.
\end{eqnarray}
Here we do not have parity symmetry for reasons that will be discussed later and consider it as a UV-complete model for RHN DM. 

\begin{table}
  \small
  \centering
  \label{tab:fields}
  \caption{Fermion and scalar contents of our model and their representations under the gauge group $SU(3)_C \times SU(2)_L \times SU(2)_R \times U(1)_{Y_L} \times U(1)_{Y_R}$.}
  \begin{tabular}{ll|ll}
  \hline\hline
  field & representation & field & representation \\
  \hline
    $Q_L=\left(\begin{matrix} u_L\\d_L \end{matrix}\right)$ &
$\left({\bf 3}, {\bf 2}, {\bf 1}, \frac{1}{3}, 0\right)$ &
  $\mathcal{Q}_{R} \ = \ \left(\begin{matrix} \mathcal{U}_R \\ \mathcal{D}_R \end{matrix}\right)$ &
  $\left({\bf 3}, {\bf 1}, {\bf 2}, 0, \frac{1}{3} \right)$ \\
  $u_R$ &
$\left({\bf 3}, {\bf 1}, {\bf 1}, \frac43, 0 \right)$ &
  $\mathcal{U}_L$ & $\left({\bf 3}, {\bf 1}, {\bf 1}, 0, \frac43 \right)$ \\
  $d_R$ &
$\left({\bf 3}, {\bf 1}, {\bf 1}, -\frac23, 0 \right)$ &
  $\mathcal{D}_L$ & $\left({\bf 3}, {\bf 1}, {\bf 1}, 0, -\frac23 \right)$ \\ \hline
    $\psi_{L}=\left(\begin{matrix} \nu_L \\ e_L \end{matrix}\right)$ &
$\left({\bf 1}, {\bf 2}, {\bf 1}, -1, 0 \right)$ &
  $\Psi_{R} = \left(\begin{matrix} N \\ \mathcal{E}_R \end{matrix}\right)$ &
  $\left({\bf 1}, {\bf 1}, {\bf 2}, 0, -1 \right)$ \\
   $e_R$ &
$\left({\bf 1}, {\bf 1}, {\bf 1}, -2, 0 \right)$ &
  $\mathcal{E}_L$ & $\left({\bf 1}, {\bf 1}, {\bf 1}, 0, -2 \right)$ \\
  \hline \hline
    $\chi_{L}=\left(\begin{matrix} \chi_L^+ \\ \chi_L^0 \end{matrix}\right)$ &
$\left({\bf 1}, {\bf 2}, {\bf 1}, 1, 0\right)$ &
  $\chi_R = \left(\begin{matrix} \chi_R^+ \\ \chi_R^0 \end{matrix}\right)$ &
  $\left({\bf 1}, {\bf 1}, {\bf 2}, 0, 1 \right)$ \\
    $\Delta_{L}=\left(\begin{matrix} \Delta_L^+/\sqrt2 & \Delta_L^{++} \\ \Delta_L^0 & -\Delta_L^+/\sqrt2 \end{matrix}\right)$ &
$\left({\bf 1}, {\bf 3}, {\bf 1}, 2, 0 \right)$ &
  $\Delta_{R} = \left(\begin{matrix} \Delta_R^+/\sqrt2 & \Delta_R^{++} \\ \Delta_R^0 & -\Delta_R^+/\sqrt2 \end{matrix}\right)$ &
  $\left({\bf 1}, {\bf 1}, {\bf 3}, 0, 2 \right)$\\ \hline
  $\Sigma_{1}$ & $\left({\bf 1}, {\bf 1}, {\bf 1}, -\frac43, \frac43 \right)$ &
  $\Sigma_{2}$ & $\left({\bf 1}, {\bf 1}, {\bf 1}, \frac23, -\frac23 \right)$ \\
  \hline\hline
  \end{tabular}
  \end{table}

The Higgs sector of the model consists of $SU(2)_L$ doublet $\chi_L$ and triplet $\Delta_L$ as well as their right-handed ``partners'' $\chi_R$ and $\Delta_R$, which are displayed in the lower part of Table~\ref{tab:fields}.
The doublets break the SM $SU(2)_L$  and the new $SU(2)_R$ gauge symmetries, as in the conventional LR models with vector-like fermions~\cite{Babu:1989rb, Berezhiani:1983hm, Rajpoot:1987fca, Davidson:1987mh, Babu:1988mw, CP1, CP2, Deppisch:2016scs}, after they obtain the non-vanishing vacuum expectation values (VEV)
\begin{eqnarray}
\langle \chi_L^0 \rangle \ = \ v_{L} \ \equiv \  v_{\rm EW} \,, \qquad
\langle \chi_R^{0} \rangle \ = \ v_R \, ,
\end{eqnarray}
with $v_{\rm EW} = 174$ GeV and $v_R$ in the TeV range or higher. The doublets also give mass to all the fermions (except the heavy and light neutrinos) through the Yukawa interactions:
\begin{eqnarray}
\label{eq:Lyukawa}
- \mathcal{L}_Y & \ \supset \ &
 y_{u}  \overline{Q}_L \widetilde{\chi}_{L} u_R
+ y_{d}\overline{Q}_L  \chi_{L} d_R
+ y_{e} \overline{\psi}_L  \chi_{L} e_R
\nonumber \\
&& +y'_{u}  \overline{\cal Q}_R \widetilde{\chi}_{R} {\cal U}_L
+ y'_{d}\overline{\cal Q}_R  \chi_{R} {\cal D}_L
+ y'_{e} \overline{\Psi}_R  \chi_{R} {\cal E}_L
+ {\rm H.c.} \,,
\end{eqnarray}
where $\widetilde{\chi}_L = i\sigma_2\chi_L^*$  and similarly for $\chi_R$ ($\sigma_2$ being the second Pauli matrix), $f_{L,\,R}$ stand for the SM chiral fermions and $\mathcal{F}_{L,\,R}$ the corresponding heavy partners. Exact parity symmetry would have implied that the Yukawa couplings $y_f = y'_f$ (and also $f = f'$ in Eq.~(\ref{eq:Lyukawa2}) below). If the $SU(2)_R\times U(1)_{Y_R}$ symmetry breaking is in the few TeV range, this would imply new quarks in the few GeV range and would be inconsistent with observations, given the current LHC bounds on vector-like quark masses in the 1--1.5 TeV range~\cite{CMS:2016usi, CMS:2016onb, ATLAS:2016ovj}. Since in our model there is no parity symmetry, the Yukawa couplings in the heavy $R$-sector are independent of those in the SM sector and  by choosing the $y^\prime$ couplings to be of order one, we can have all the heavy fermions close to or above the TeV scale if $v_R \gtrsim  $ few TeV. It is worth noting that at this stage unlike the standard minimal LR seesaw model, the lower bound on $M_{W_R}$~\cite{Kkbar1, Kkbar2, Kkbar3} from low-energy flavor changing effects such as $K^0-\overline{K}^0$ mixing does not apply to our model.


The triplet scalars $\Delta_{L,\,R}$ are used to generate the masses of light and heavy neutrinos via type-II seesaw mechanism~\cite{type2a, type2b, type2c, type2d}. To get the VEVs for the triplets, we choose the Higgs potential of the form:
\begin{eqnarray}
\label{eqn:potential}
\mathcal{V} & \ = \ &
- \mu^2_{L} (\chi^\dagger_L \chi_L)
-{\mu}^2_{R} (\chi^{\dagger}_R \chi_R)
+ M^2_L {\rm Tr} ( \Delta^\dagger_L \Delta_L )
+ M_R^2 {\rm Tr} ( {\Delta}^\dagger_R \Delta_R ) \nonumber \\
&& + m_{L} \chi_L^T i\sigma_2 \Delta_L^\dagger \chi_L
   + m_{R} \chi_R^T i\sigma_2 \Delta_R^\dagger \chi_R
+ {\rm H.c.} \, .
\end{eqnarray}
In the above expression, we have omitted the quartic terms of the form $(\chi^\dagger\chi)^2$ etc. since they do not affect our results and shown only the terms relevant for heavy and light neutrino masses after spontaneous symmetry breaking:
\begin{eqnarray}
\langle \Delta^0_L \rangle \ = \  w_L  \ \sim \ \frac{m_L v^2_{\rm EW}}{M_L^2} \,, \qquad
\langle \Delta^0_R \rangle \ = \ w_R \ \sim \ \frac{m_R v_R^{2}}{M_R^ 2} \,.
\label{eq:2.8}
\end{eqnarray}
We choose the parameters of the model such that $w_L\sim $ eV (corresponding to the parameter $m_L \sim 10^{-6}$ GeV) and  $w_R \sim v_R \sim 10$ TeV. Given the Yukawa interactions
\begin{eqnarray}
\label{eq:Lyukawa2}
- \mathcal{L}_Y & \ \supset \ &
f \overline{ \psi}_L^C i\sigma_2 \Delta_L \psi_L 
+ f' \overline{ \Psi}_R^C i\sigma_2 \Delta_R \Psi_R
+ {\rm H.c.} \,,
\end{eqnarray}
the $\Delta_R$ term gives masses to the RH neutrinos of order of few to 10 TeV whereas $w_L$ gives masses to the left handed neutrinos via the type-II seesaw mechanism. In the LR scenario without parity at the TeV scale, the $f$ and $f'$ couplings are independent parameters, i.e. $f \neq f'$, thus the RHN sector and the DM phenomenology in this paper are completely independent of the light neutrino sector. There are enough free parameters in the $f$-couplings that all neutrino oscillation parameters can be fitted without modifying the DM phenomenology discussed here. The $f$ and $f'$ couplings however could be made equal in parity symmetric models, with slight extension of the Higgs sector; see Section~\ref{sec:2.2}.

It is remarkable that at the level of dimension four interactions, i.e.~the Yukawa and gauge interactions and ignoring non-perturbative effects, the model has a large global symmetry: $U(1)_{B,\,L}\times U(1)_{B,\,R}\times Z_{2\ell,\,L} \times Z_{2 \ell,\, R}$ even after the symmetry breaking VEVs are turned on. Here $U(1)_{B,L}$ is defined as the baryon number of SM quark fields $Q,u,d$ and $U(1)_{B,R}$ the baryon number of the heavy quark fields of the $SU(2)_R$ sector. In the leptonic sector, the residual symmetries are two discrete symmetries, i.e.~$Z_{2\ell,\,L}$ defined as $(-1)^{n_\ell}$ where $n_\ell$ is the lepton number of the SM leptons and similarly $Z_{2\ell,\,R}$ for the heavy leptons of the $SU(2)_R$ sector. One of the most important implications of these symmetries is that the lightest of the heavy baryons made out of $QQQ, QQq,Qqq$ is absolutely stable and the lightest of the heavy leptons is also absolutely stable.
We will see later in Section~\ref{sec:2.5} how these heavy baryons can be depleted during the cosmological evolution by introduction of new terms in the Lagrangian. We further note that if we choose the lightest lepton of the heavy sector to be the lightest of the RHNs, it will remain absolutely stable and can play the role of cold DM of the Universe. In the rest of the paper, we study the phenomenological implications of this model for DM and collider signals.

For the phenomenological purpose of avoiding the heavy lightest baryons also becoming DM and affecting the evolution of the Universe, we add two more Higgs singlets $\Sigma_{1,2}$, as shown in Table~\ref{tab:fields}. Both these fields have non-vanishing VEVs and connect the heavy and light singlet quarks via the following terms:\footnote{These multiplets are similar to those introduced in Ref.~\cite{gu}. However, unlike in Ref.~\cite{gu}, we do not have the leptophilic scalar $\Sigma_E$, so that the $Z_{2\ell,\,L}\times Z_{2\ell,\,R}$ symmetry remains exact. }
\begin{eqnarray}
\label{eqn:Lmix}
{\cal L}_{\rm mix} \ = \ \lambda_U \overline{\mathcal{U}}_Lu_R\Sigma^*_1+\lambda_D\overline{\mathcal{D}}_Ld_R\Sigma^*_2+ {\rm H.c.} \,.
\end{eqnarray}
Once the $\Sigma$ fields acquire VEVs, new heavy-light quark mass mixing terms $\delta_{U} = \lambda_{U} \langle \Sigma_1 \rangle$ and $\delta_D = \lambda_D \langle \Sigma_2 \rangle$ appear in the Lagrangian which break the global symmetry down to $U(1)_{B} \times Z_{2{\ell},\,L}\times Z_{2{\ell},\, R}$ at the tree level. Note that the $Z_2$ symmetry responsible for the stability of RHN DM is a subgroup of this symmetry under which the $L$-sector leptons are even and the $R$-sector leptons are odd, and thus it is a symmetry of the theory that remains at the renormalizable level.\footnote{Note that if non-renormalizable terms such as $\overline{\cal E}_Le_R\Sigma^2_1\Sigma_2/\Lambda^2$ terms are included, this symmetry will be broken and will make the RHN DM unstable; however a proper choice of $\Lambda$ will make the DM long lived~\cite{dkmtz}. We do not discuss this possibility here.}  These singlet VEVs also break the $U(1)_{Y_L} \times U(1)_{Y_R}$ gauge symmetries down to $U(1)_{\mathcal Y}$ below the $\langle \Sigma \rangle$ scale. This is the version of the model discussed in Ref.~\cite{bem,dkmtz}. We note that $\mathcal Y$ quantum numbers are different from the SM hypercharge (which we denote as usual by $Y$) and $B-L$ of the LR models.

\subsection{Parity symmetric version of the model} \label{sec:2.2}
The minimal model discussed above cannot be made parity symmetric for RH scale in the 1--10 TeV range. The reason is that parity symmetry would imply the Yukawa couplings in the two sectors to be equal. As a result, the lightest heavy quark mass in the 1 GeV range makes the minimal parity conserving theory phenomenologically untenable. The lowest right handed scale that would make this minimal theory acceptable is $v_R \sim 10^7-10^8$ GeV.

A simple extension of the minimal model that can make it a viable theory for ${\cal O}$(10 TeV) RH scale, is to double the number of electroweak doublets $\chi_{L,R}$ in both sectors. The resulting Yukawa couplings for this case can be written as
\begin{eqnarray}
\label{eq:Lyukawap}
- \mathcal{L}_Y & \ \supset \ &
 \sum_{a=1,2} \left[ y_{u,a}  \overline{Q}_L \tilde{\chi}_{L,a} u_R
+ y_{d,a}\overline{Q}_L  \chi_{L,a} d_R
+ y_{e,a} \overline{\psi}_L  \chi_{L,a} e_R \right]
\nonumber \\
&& + 
(f_{L,\,R} \to \mathcal{F}_{L,\,R},\, \chi_{L,a} \to \chi_{R,a}
)
+ {\rm H.c.} \,.
\end{eqnarray}
Note that the Yukawa couplings in the $L$ and $R$ sectors are now equal due to parity symmetry, unlike in Eq.~\eqref{eq:Lyukawa}.
We then arrange the soft breaking mass terms for the $\chi_{L,R}$ so that $\langle \chi_{L,1} \rangle \simeq 0$ and  $\langle \chi_{L,2} \rangle = v_{\rm EW}$ with $y_{u,d,e;1}\simeq 1$ whereas
$y_{u,2}:y_{c,2}:y_{t,2} ~=~ m_u : m_c: m_t$ and similarly for the down-type quark and charged lepton Yukawa couplings $y_{f,2}$. In the RH sector, if we choose the VEVs of $\langle \chi_{R,a} \rangle \sim v_R\simeq 1-10$ TeV, then the $R$-sector quark and lepton masses come almost entirely from the $y_{u,d,e;2}$ couplings, and as a result, the $R$-sector heavy quark and lepton masses are in the TeV range as required by current LHC limits~\cite{CMS:2016usi, CMS:2016onb, ATLAS:2016ovj}. To give masses to the light and heavy neutrinos, we introduce only one triplet $\Delta_{L,R}$ with $Y_{L,R}=2$ as in the parity broken model discussed in Section~\ref{sec:2.1}. Parity symmetry now makes the masses in the two sectors proportional and mixings equal. Similar models with exact parity relating the heavy and light sectors were considered in Refs.~\cite{Babu:1989rb, barr, hook} for the strong CP problem. 

In what follows, we only consider the minimal parity broken model of Section~\ref{sec:2.1}.

\subsection{Effective theory at the $SU(2)_L \times SU(2)_R \times U(1)_{\mathcal Y}$  level} \label{sec:2.3}

In discussions of the phenomenological consequences below, the presence of higher gauge symmetry  $SU(2)_L \times SU(2)_R \times U(1)_{Y_L} \times U(1)_{Y_R}$ is not important, but rather the reduced  symmetry $SU(2)_L \times SU(2)_R \times U(1)_{\mathcal Y}$ that emerges after the $\Sigma$-like fields acquire VEVs. Though the $\mathcal Y$ numbers of SM fermions are the same as the SM hypercharges, $U(1)_{\mathcal Y}$ is not the SM gauge group $U(1)_Y$, with the latter a combination of the former and the $U(1)$ subgroup of $SU(2)_R$ after symmetry breaking. At this level, the gauge interactions become essentially that of LR models with quark seesaw~\cite{Babu:1989rb, CP1, CP2}, whereas the Yukawa interactions become different due to lack of parity symmetry. In addition, there are heavy and light quark mixings induced by
\begin{eqnarray}
\label{eq:Lmix2}
{\cal L}_{\rm mix} \ = \ \delta_U \overline{\mathcal{U}}_Lu_R+\delta_D\overline{\mathcal{D}}_Ld_R+ {\rm H.c.} \,,
\end{eqnarray}
which arises from the Lagrangian in Eq.~(\ref{eqn:Lmix}). The global symmetries discussed above remain valid at this level. We assume that the heavy gauge boson $Z_{LR}$ which emerges from the breaking of $U(1)_{Y_L} \times U(1)_{Y_R} \to U(1)_{\mathcal{Y}}$ has decoupled. Thus in the neutral gauge boson sector we can consider only the mixing involving the SM $Z$ boson and the conventional $Z_R$ boson in LR models (see Section~\ref{sec:2.4} below), which is similar as in Refs.~\cite{Babu:1989rb, CP1, CP2} except that in our model the triplet $\Delta_R$ also contributes to $Z_R$ and $W_R$ masses.

\subsection{Mixing between the heavy and light sectors}\label{sec:2.4}

In this subsection we illustrate explicitly how the RHN DM particle in the heavy sector interacts with the SM fields, which will pave the way for all the phenomenological discussions below on the DM  annihilation in the early Universe, its direct and indirect detection, and collider searches.

In absence of any Higgs and fermion mixings connecting the heavy and light sectors, the RHN DM could only interact with the SM fermions through the heavy $Z_R$ boson, which couples directly to the SM quarks and leptons through the $U(1)_{\mathcal Y}$ interaction. 
Once the mixing terms are included, in the lowest order, the RHN DM could talk to the SM fields via the scalar mixing ($h - \Delta_R^0$)  and the neutral gauge boson mixing ($Z - Z_R$) at the tree level. Regarding the physical scalars, only two are directly relevant to the DM phenomenology, i.e. the SM Higgs $h$ and the neutral CP-even $\Delta_R^0$ from the triplet $\Delta_R$. The SM Higgs $h$ is assumed to be predominantly from the doublet $\chi_{L}$, as in the conventional LR models~\cite{Babu:1989rb, CP1, CP2}, while $\Delta_R^0$ couples directly to the RHNs. The mixing of the two scalars could be induced from the quartic couplings of form $\lambda ( \chi_L^\dagger \chi_L ) {\rm Tr} ( \Delta_R^{\dagger} \Delta_R )$. Here for simplicity we have neglected the effects from other scalars. After spontaneous symmetry breaking, the $h - \Delta_R^0$ mixing reads 
\begin{eqnarray}
\label{eqn:mixs}
\zeta_S \ \simeq \ \frac{\lambda v_{\rm EW}}{w_R} \, ,
\end{eqnarray}
with $w_R$ the non-vanishing VEV of the RH triplet $\Delta_R$ as defined in Eq.~\eqref{eq:2.8}. Note that the scalar mixing $\zeta_S$ contributes to the SM Higgs mass square, at the order of $- (\lambda v_{\rm EW} w_R )^2 / w_R^2 = - \lambda^2 v_{\rm EW}^2$, which requires a larger quartic coupling than in SM and thus could possibly help to improve the stability of the EW vacuum, as compared to the SM~\cite{CP1}.


Regarding the $Z - Z_R$ mixing, the neutral gauge boson mass matrix reads as follows, after symmetry breaking at the RH scale and the EW scale, in the basis of ($W_{3L}$, $W_{3R}$, $B$) with $B$ the gauge boson for the $U(1)_{\mathcal Y}$ symmetry:
\begin{eqnarray}
{\cal M}_{\rm neutral}^2 \ = \ \left( \begin{matrix}
\frac12 g_L^2 (v_{\rm EW}^2 + 4w_L^2) & 0 & -\frac12 g_L g_{\mathcal Y} (v_{\rm EW}^2 + 4w_L^2) \\
0 & \frac12 g_R^2 (v_R^2 + 4w_R^2) & -\frac12 g_R g_{\mathcal Y} (v_R^2 + 4w_R^2) \\
-\frac12 g_L g_{\mathcal Y} (v_{\rm EW}^2 + 4w_L^2) & -\frac12 g_R g_{\mathcal Y} (v_R^2 + 4w_R^2) & \frac12 g_{\mathcal Y}^2 (v_{\rm EW}^2 + v_R^2 + 4w_L^2 + 4w_R^2)
\end{matrix} \right) \,,\nonumber \\ \label{mneusq}
\end{eqnarray}
where $g_L$, $g_R$ and $g_{\cal Y}$ are the $SU(2)_L$, $SU(2)_R$ and $U(1)_{\cal Y}$ gauge couplings, respectively.
The VEVs in the mass matrix~\eqref{mneusq} have the hierarchical structure $w_L \ll v_{\rm EW} \ll v_R, \, w_R$, and in this limit, the matrix can be diagonalized by the unitarity rotation
\begin{eqnarray}
\label{eqn:gaugemix}
\left(\begin{array}{ccc}
W_H^3 \\ W_Z^3 \\ A
\end{array} \right) \ = \
\left(\begin{array}{ccc}
0 & \cos\phi & -\sin\phi \\
\cos\theta_W & - \sin\theta_W \sin\phi & - \sin\theta_W \cos\phi \\
\sin\theta_W & \cos\theta_W \sin\phi &  \cos\theta_W \cos\phi \\
\end{array} \right)
\left(\begin{array}{ccc}
W_{3L} \\ W_{3R} \\ B
\end{array} \right) \,,
\end{eqnarray}
where $\theta_W$ is the Weinberg angle and $\tan\phi \equiv g_{\mathcal Y} / g_R$. Obviously $A$ is the massless photon, and we are left with two massive states, i.e. the SM $Z$ boson and the heavy $Z_R$ boson, with masses respectively given by
\begin{eqnarray}
M_Z^2 & \ = \ & \frac{g_L^2}{2 \cos^2\theta_W} v^2_{\rm EW} \,, \nonumber \\
M_{Z_R}^2 & \ = \ & \frac12 (g_R^2 + g_{\mathcal Y}^2) ( v_R^2 + 4w_R^2 ) \,,
\end{eqnarray}
and the mixing angle at the leading order given by
\begin{eqnarray}
\label{eqn:mixz}
\zeta_Z \ \simeq \
\frac{\sin^3\phi\cos\phi}{\sin\theta_W} \frac{v_{\rm EW}^2}{v_R^2 +4w_R^2}
\ \equiv \ \xi_Z \left( \frac{M_Z}{M_{Z_R}} \right)^2 \, ,
\end{eqnarray}
where the parameter
\begin{eqnarray}
\xi_Z \ = \ \sin\theta_W \left[ \frac{g_R^2}{g_L^2} \cot^2\theta_W - 1 \right]^{-1/2} \,.
\end{eqnarray}
To zeroth order in $\xi_Z$, the $Z$ couplings are the same as in the SM, as they should be. There are small corrections of order $\xi_Z\ll 1$.
It should be noted that when the gauge coupling $g_R$ approaches the theoretical lower limit $g_L \tan\theta_W$ (which is independent of the symmetry breaking pattern~\cite{Dev:2016dja}), the $\xi_Z$ parameter, and therefore the $Z - Z_R$ mixing, could be significantly enhanced.

\begin{figure}[t!]
  \centering
   \begin{tikzpicture}[thick]
  \draw[ZZ](-3,0)--(-4.5,0)node[left]{$W$};
  \draw[ZZ](3,0)--(4.5,0)node[right]{$W_R$};
  \draw[quark] (3,0)  arc (0:55 : 3 and 1.5) ;
  \draw[antiheavyquark] (0,1.5)  arc (90:55 : 3 and 1.5);
  \draw[heavyquark] (0,1.5)  arc (90:125 : 3 and 1.5);
  \draw[antiquark] (-3,0)  arc (180:125 : 3 and 1.5);
  \draw (0.3,1.5) node [rotate=0] {$\times$};
  \draw (0.3,-1.5) node [rotate=0] {$\times$};
  \draw (1.88,1.48) node [rotate=65] {$\times$};
  \draw (1.88,-1.48) node [rotate=-65] {$\times$};
  \draw (-1.88,1.48) node [rotate=115] {$\times$};
  \draw (-1.88,-1.48) node [rotate=-115] {$\times$};
  \draw[quark] (-3,0)  arc (180:235 : 3 and 1.5);
  \draw[antiheavyquark] (0,-1.5)  arc (270:235 : 3 and 1.5);
  \draw[heavyquark] (0,-1.5)  arc (270:305 : 3 and 1.5);
  \draw[antiquark] (3,0)  arc (360:305 : 3 and 1.5);
  \draw (-2.25,1.15) node [rotate=0] {$t_L$};
  \draw (-2.25,-1.1) node [rotate=0] {$b_L$};
  \draw (2.9,1.2) node [rotate=0] {$T_R$};
  \draw (2.9,-1.25) node [rotate=0] {$B_R$};
  \draw (-0.5,1.75) node [rotate=0] {$t_R$};
  \draw (-0.5,-1.75) node [rotate=0] {$b_R$};
  \draw (1.4,1.75) node [rotate=0] {$T_L$};
  \draw (1.4,-1.75) node [rotate=0] {$B_L$};
  \draw (0.4,1.1) node [rotate=0] {$\delta_U$};
  \draw (0.4,-1.1) node [rotate=0] {$\delta_D$};
  \draw[color=white](0,-1.4)->(1,-1.4);
  \end{tikzpicture}
  \caption{Feynman diagram for the 1-loop $W - W_R$ mixing induced by the $\delta_{U,\,D}$ terms.}
  \label{fig:Wmixing}
\end{figure}
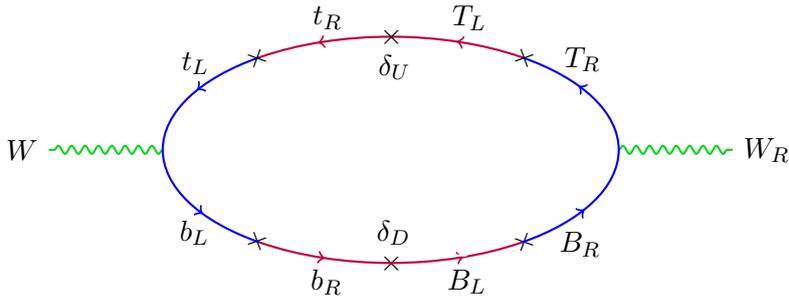

In the minimal version of our model, we do not have the bi-multiplet scalars which transform non-trivially under both $SU(2)_L$ and $SU(2)_R$, like the bi-doublet $({\bf 1}, {\bf 2}, {\bf 2}, 0)$ in the conventional LR models; thus the charged gauge bosons $W$ and $W_R$ can not mix at the tree level. Only when the SM and heavy fermions talk to each other via the mixing terms in Eq.~(\ref{eq:Lmix2}), can the $W - W_R$ mixing be generated at the 1-loop level. The largest contribution stems from the mixing of third generation quarks and their heavy partners, and the corresponding Feynman diagram is presented in Fig.~\ref{fig:Wmixing}. At the leading order the charged gauge boson mixing parameter reads
\begin{eqnarray}
\label{eqn:Wmixing}
\zeta_{W} \ \simeq \
\frac{g_L g_R \delta_U \delta_D m_t m_b}
{16\pi^2 M_T M_B M^2_{W_R}} \,,
\end{eqnarray}
with $m_{t,\,b}$ the masses of SM top and bottom quarks, and $M_{T,\,B}$ the masses of heavy top and bottom partner fermions, respectively. Since the $\delta_{U,\,D}$ terms break the global symmetry $U(1)_{B,\,L}\times U(1)_{B,\,R}$, they are expected to be small, and therefore, the induced $W-W_R$ mixing is also a small number: for instance, $\zeta_{W} \sim 10^{-12}\delta_U\delta_D$ GeV$^{-2}$ for TeV-scale heavy partners.

The $W_L-W_R$ mixing and the mixing between light and heavy quarks are crucial to deplete the heavy hadronic states (see Section~\ref{sec:2.5} below). One point to note is that these mixings do not affect the stability of the lightest heavy lepton, as it is odd under the $Z_{2 \ell}$ symmetry, and can be the DM candidate.



\subsection{Depleting the heavy hadronic states} \label{sec:2.5}
This model has two classes of baryons and mesons: (i) light baryons $(qqq)$ and mesons $(\bar{q}q$)  that are part of the SM and (ii) heavy baryons and mesons which arise due to the fact that QCD is shared by the heavy quarks. In this subsection, we focus on the heavy baryons and mesons, and discuss the constraints on their properties from cosmology.  Let us first note that there will be three kinds of baryons $(QQQ, QQq, Qqq)$ and two kinds of mesons $(\bar{Q}q, \bar{Q}Q)$ involving heavy quarks. In the absence of any mixing between the heavy and light quark sector, the lightest of all these five kinds of states will be stable and their abundance in the early Universe will be determined by their masses, as in the case of strongly interacting DM~\cite{Goldberg:1986nk, Rich:1987st, Chivukula:1989cc, Starkman:1990nj, Nardi:1990ku, nussinov}. Rough estimates in Ref.~\cite{nussinov} give that $\Omega_Q/\Omega_B\sim 10 \: M_Q/m_p$ where $M_Q$ stands for the mass of lightest baryonic or mesonic state involving the heavy quark $Q$ and $m_p$ is the mass of proton. This means that stable dark baryons or mesons above the GeV scale already over-close the Universe. Our goal is however to have the RHN as the only DM. Furthermore, heavy colored particles with masses up to a few TeV are incompatible with bounds on anomalous nuclei~\cite{Hemmick:1989ns, MT, Mohapatra:1999gg}. Masses heavier than this also seem to be excluded by considerations of DM-cosmic ray interactions producing gamma rays~\cite{Wandelt:2000ad} and DM capture and self-annihilation in Earth's core producing internal heat flow~\cite{Mack:2007xj, Mack:2012ju}. So we would like to provide a mechanism to deplete the heavy quark bound states. The simplest way to do that is to introduce the heavy-light quark mixings given in Eq.~(\ref{eqn:Lmix}). In fact requiring that they are depleted by the time of QCD phase transition temperature $T_{\rm QCD}\approx 200$ MeV imposes {\it lower} bounds on the magnitudes of $\delta_{U,\,D}$ which we estimate below.

To calculate the lower limits on the $\delta_{U,\,D}$, we note that they generate heavy-light quark mixings which can be determined by analyzing the following mass matrix, in the basis of $(q_R, Q_R)$ to $(q_L, Q_L)$,
\begin{eqnarray}
M_{qQ}~=~\left(\begin{matrix} yv_L & \delta_Q\\0 & y'v_R\end{matrix}\right) \,.
\end{eqnarray}
Diagonalizing this mass matrix, one can find that the heavy-light quark mixing in the right-handed sector is given by $\beta_R \simeq \delta_Q / y'v_R$ and in the left-handed sector by $\beta_L \simeq yv_L\delta_Q / y^{\prime 2}v^2_R$. Thus the RH sector mixing is expected to be much larger. This mixing will allow states with heavy quarks to decay to light states plus SM gauge bosons. Typically  for the lightest meson in the heavy sector, the decay goes like $\bar{Q}q\to \bar{q}q+ Z$. This decay process is much like the decay of free heavy vector-like quarks. The decay width can be estimated to be
\begin{eqnarray}
\Gamma_{Q} \ \simeq \ \frac{g_L^2}{64\pi \cos^2\theta_W} \frac{\delta^2_Q}{v_R^2} \frac{M_Q^3}{M_Z^2}\, .
\label{MQq}
\end{eqnarray}
where $M_Q$ is the mass of the heavy vector-like quark with $M_{Q}\gg M_Z$.
Equating the decay rate~\eqref{MQq} to the Hubble rate $1.66\: g_*^{1/2}T^2/M_{\rm Pl}$ (where $g_*$ is the effective relativistic degrees of freedom at temperature $T$ and $M_{\rm Pl}$ is the Planck mass) and taking the decay temperature $T\sim 1$ GeV, we get a lower limit on the heavy-light quark mixing parameter, $\delta_Q \gtrsim 10^{-6}$ GeV above which value the heavy mesons will decay before QCD phase transition and not survive as the cold DM of the Universe. For the baryons, the decay width depends on the number of heavy quarks via $(\delta_Q/v_R)^{2N_Q}$. Thus, it follows that if we choose the value of $\delta_Q \gtrsim 10^{-6}$ GeV, all heavy hadrons will disappear from the Universe before QCD phase transition, thus preserving the success of the Big Bang Nucleosynthesis.

\section{DM relic density}
\label{sec:density}


In order to determine the relic density of RHN DM, we note that in the early Universe, all the particles were in equilibrium with the light SM sector particles due to the common $SU(3)_C$ color and $U(1)_{\mathcal Y}$ interactions. As the Universe cools, the particles of the heavy sector being heavier than the DM $N$, slowly annihilate away leaving the $N$'s in the primordial plasma. As the temperature falls below $M_N$, the DM density goes down and freezes out for $T\lesssim M_N/20$, as in case of a generic cold DM candidate. The primary annihilation channels to the SM particles proceed via particles that connect the two sectors, i.e. the SM $Z$ boson and the heavy $Z_R$ boson which mix at the tree level, as well as the $h - \Delta_R^0$ Higgs portal. Both the $Z$ and Higgs portals are suppressed by the small mixing angles $\zeta_Z$ and $\zeta_S$ connecting the light and heavy sector, which are respectively of order $\lambda v_{\rm EW} / v_R$ and $v_{\rm EW}^2 / v_R^2$. On the other hand, though the couplings of $Z_R$ to $N$ and the SM fermions are of order one, the $Z_R$ portal is however suppressed by the large $Z_R$ mass, except near the resonance $2M_N \simeq M_{Z_R}$. Since the $Z_R$ channel is very important for the RHN DM relic density calculation, we will first present the current limits on $Z_R$ mass in our model from direct collider searches.

\subsection{Limits on $Z_R$ mass} \label{sec:3.1}
A heavy $Z'$ boson could decay promptly into the SM quarks and leptons, and the $\sqrt s=13$ TeV LHC dilepton searches require that $M_{Z'} \gtrsim 3$ TeV~\cite{Aaboud:2016cth} for a sequential $Z'$ model with SM-like couplings~\cite{Langacker:2008yv}. However, the exact limit in LR models depends on the specific value of $g_R/g_L$~\cite{Patra:2015bga, Lindner:2016lpp}, and moreover, in our case, the corresponding limits will be slightly weaker than in conventional LR models if we have the additional decay mode $Z_R\to NN$ kinematically open. In this subsection, we reinterpret the $\sqrt s=13$ TeV LHC limits on $Z'$ mass for our $Z_R$ scenario.

\begin{figure}[t!]
  \centering
  \includegraphics[width=0.6\textwidth]{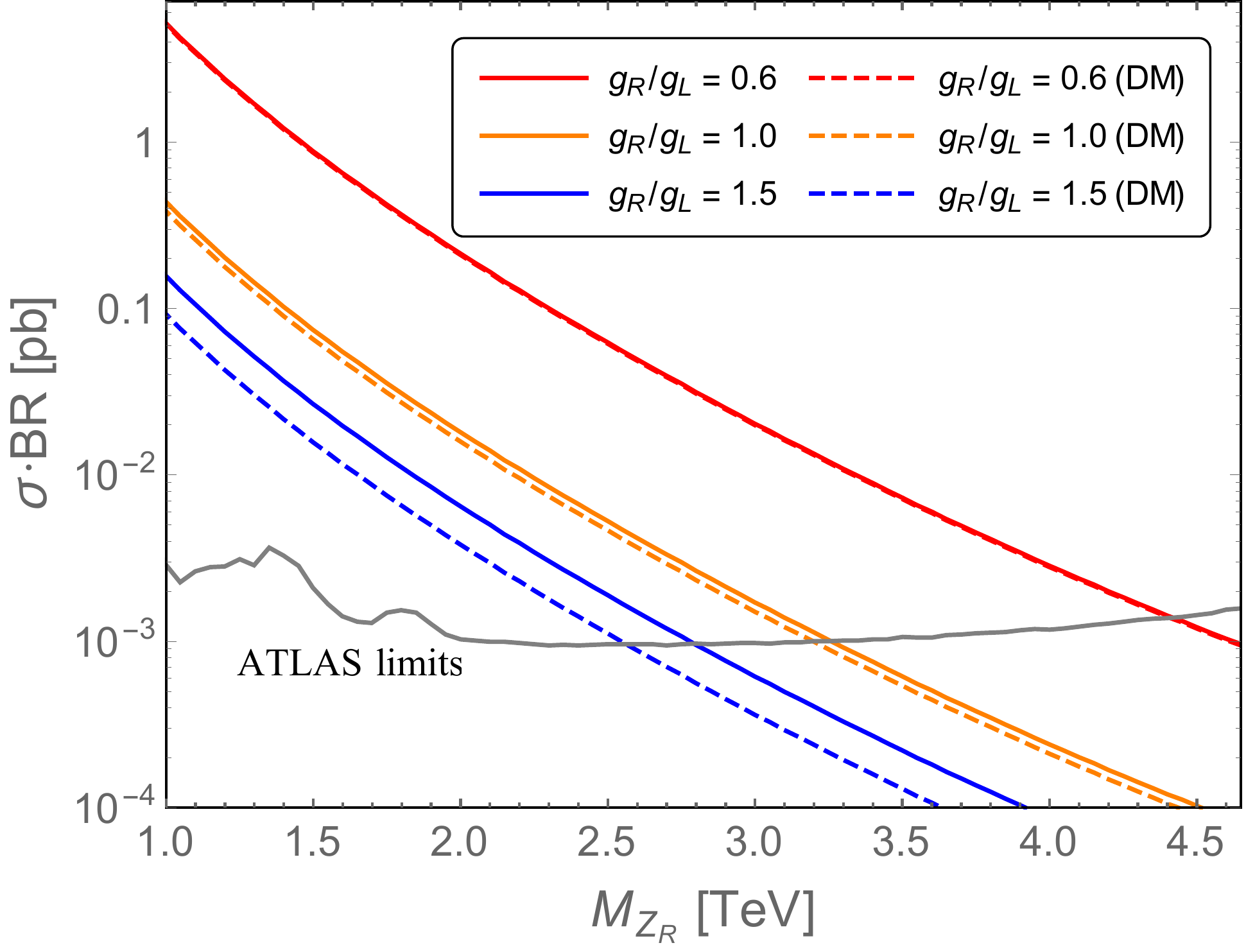}
  \caption{Model predictions for the $Z_R$ production cross section times branching ratio to dileptons at the $\sqrt s=13$ TeV LHC for three different values of $g_R/g_L$. The solid (dashed) curves are without (with) including the $Z_R$ decay to $NN$ for its total width. The gray curve shows the current 95\% CL upper limit from ATLAS~\cite{Aaboud:2016cth}.}
\label{fig:zprime}
\end{figure}

To do this, we use the heavy neutral boson $Z'_{\rm SSM}$ in sequential SM~\cite{Langacker:2008yv} as a benchmark scenario, and rescale the cross section times branching ratio to electrons and muons and the corresponding mass limit as reported in Ref.~\cite{Aaboud:2016cth}. In our model, the couplings of $Z_R$ to the SM fermions, as well as to the DM, are essentially proportional to their quantum numbers $\mathcal{Y} = Y_L + Y_R$, rescaled by the gauge mixing $\sin\phi$; cf. the rotation matrix in Eq.~(\ref{eqn:gaugemix}). For the gauge coupling $g_R = g_L$ in our model, the $Z_R$ production rate is 5 times smaller than a $Z'_{\rm SSM}$; on the other hand, the branching ratio to $e^+ e^-$ and $\mu^+ \mu^-$ is $1/4$, which is much larger than that of $Z'_{\rm SSM}$, if the decay mode of $Z_R$ to DM pair is not kinematically allowed. Then in our model the cross section times branching ratio of $Z_R$ is 0.80 times that of a $Z'_{\rm SSM}$ boson. Including the $Z_R\to NN$ decay mode, this limit goes down slightly. When $g_{R}$ is different from $g_L$, due to the gauge coupling dependence of $Z_R$ production rate (as well as the lepton branching ratio if the DM mode is open), the $Z_R$ mass limits will be weaker (stronger) if $g_R$ becomes smaller (larger) than $g_L$. For the sake of comparison, we show in Fig.~\ref{fig:zprime} our model predictions for the $Z_R$ production cross section times dilepton branching ratio as a function of the $Z_R$ mass for three representative values of $g_R/g_L$ and compare it with the latest 95\% CL from ATLAS~\cite{Aaboud:2016cth} to derive the bounds on $M_{Z_R}$ and $w_R$, as listed in Table~\ref{tab:cs}.

\begin{table}[t!]
  \centering
  \caption[]{The lower limit on $M_{Z_R}$ and the corresponding RH VEV $w_R$ (assuming $v_R = w_R$) from LHC dilepton constraints~\cite{Aaboud:2016cth}. The values in brackets assume that the decay mode $Z_R \to NN$ is also kinematically allowed and $M_N \ll M_{Z_R}$.}
  \label{tab:cs}
  \begin{tabular}{ccc}
  \hline\hline
  $g_R / g_L$ & $M_{Z_R}$ [TeV] & $w_R$ [TeV] \\ \hline
  0.6 & 4.4 (4.4) & 2.9 (2.9)\\
  1.0 &  3.3 (3.2) & 2.6 (2.6) \\
  1.5 &  2.8 (2.6) & 1.7 (1.5) \\
  \hline\hline
  \end{tabular}
\end{table}



\subsection{Dominant annihilation channels} \label{sec:3.2}
For completeness, we will consider both the scalar and gauge boson mediated channels, i.e.
\begin{eqnarray}
&& NN \ \to \ h^{(\ast)} / \Delta_R^{0 \, (\ast)} \ \to \  \text{SM particles} \,, \nonumber \\
&& NN \ \to \ Z^{(\ast)} / Z_R^{(\ast)} \ \to \ \text{SM particles} \,.
\label{eq:ann}
\end{eqnarray}
Note that in current model the SM Higgs $h$ and $\Delta_R^0$ could in principle mix with the other scalars from the singlets $\Sigma_{1,\,2}$, the doublet $\chi_R$ and the triplet $\Delta_L$, and these scalars also contribute to the annihilation of $N$. However, we can always choose the quartic couplings in the scalar potential such that the mixings of $h$ and $\Delta_R^0$ to these scalars are negligible.
The thermally averaged DM annihilation cross section times velocity $\langle \sigma v \rangle$ in various channels are collected in Appendix~\ref{app:A}, where we list explicitly the coefficients $a$ and $b$ in Taylor expansion $\langle \sigma v \rangle = a + b \langle v^2 \rangle + \mathcal{O} (v^4)$. Combining all these channels, we find that the annihilation of DM in our model depends on the gauge coupling $g_R$, quartic coupling $\lambda$, as well as the RH scale $w_R$ and the masses and widths of $\Delta_R^0$ and $Z_R$, in addition to DM mass $M_N$. The key Yukawa coupling $f'$ is related to the DM mass via the relation $M_N = 2 f' w_R$, which implies that for a light DM with $f' = M_N / 2 w_R \ll 1$ the scalar portal is further suppressed by the small Yukawa coupling $f'$.

\begin{figure}[t!]
  \centering
  \includegraphics[width=0.6\textwidth]{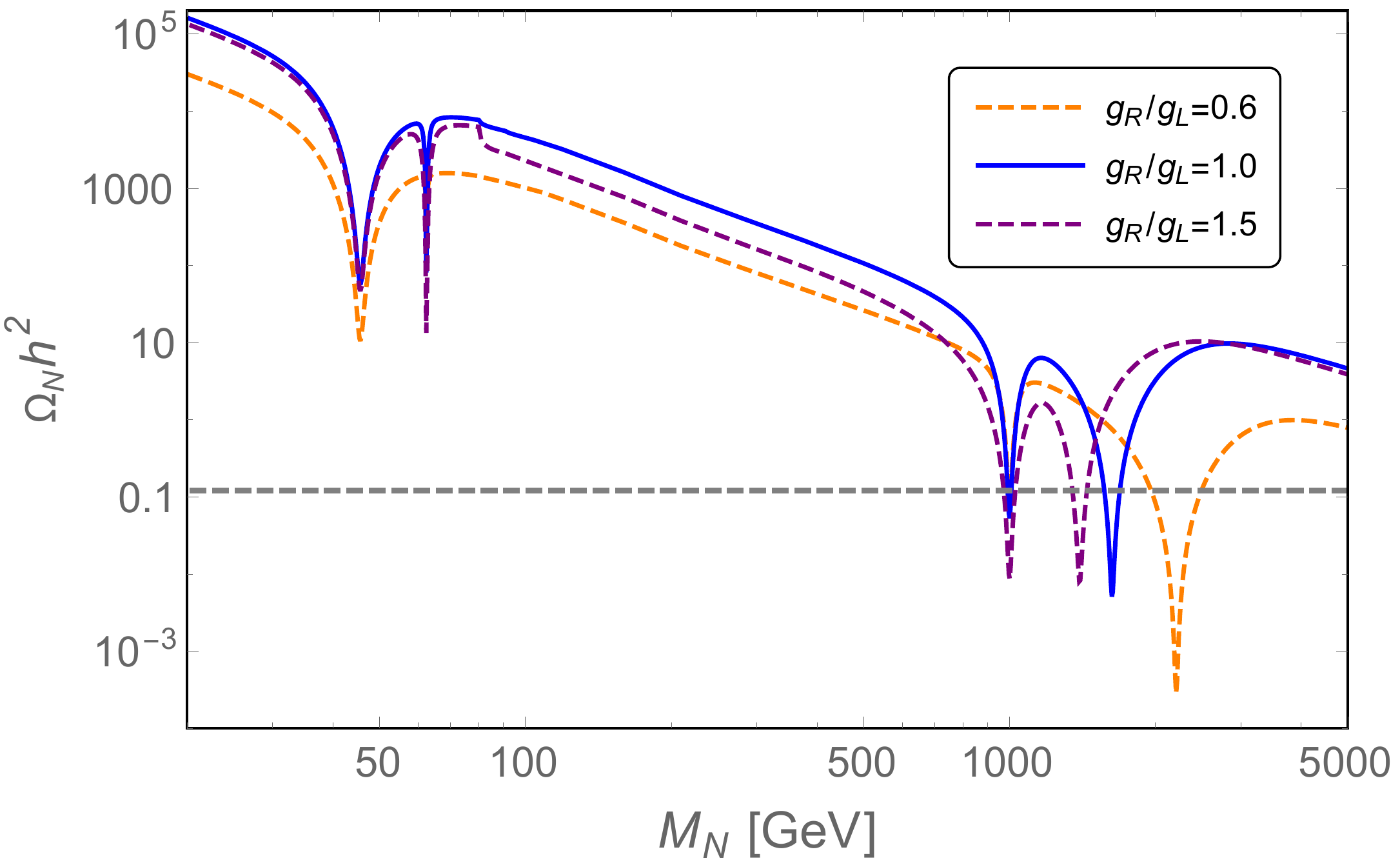} \\
  \caption{An illustration of the relic density of the RHN DM as a function of its mass $M_N$ for different values of  $g_R/g_L$. The horizontal line gives the observed value from Planck data~\cite{Ade:2015xua}. See text for details of the model parameters chosen here. }
  \label{fig:density}
\end{figure}

Once the coefficients $a$ and $b$ are known for all the available channels, the relic density of the DM can be calculated using the general formula~\cite{Kolb:1990vq}
\begin{eqnarray}
\Omega_N h^2 = \frac{1.07 \times 10^9 \, {\rm GeV}^{-1}}{M_{\rm Pl}} \frac{x_F}{\sqrt{g_\ast}} \frac{1}{a+3b/x_F} \,,
\label{eq:relic1}
\end{eqnarray}
where $x_F = M_N / T_F \simeq 20$ (with $T_F$ being the freeze-out temperature), $g_\ast = 106.75$ the relativistic degrees of freedom at $T_F$, and $a$ and $b$ the annihilation coefficients summing up all the available channels. An example is given in Fig.~\ref{fig:density}, where we set explicitly the quartic coupling $\lambda = 1$ and the heavy scalar masses $M_{\Delta_R^0} = 2$ TeV with width 30 GeV. Three different value of the gauge coupling $g_R$ are chosen: $g_R / g_L = 0.6$, 1 and 1.5. The corresponding $Z_R$ mass and the RH scale $v_R = w_R$ are set to their current experimental constraints listed in Table~\ref{tab:cs} without the $Z_R\to NN$ decay mode. The horizontal dashed line shows the observed relic density, as measured by Planck~\cite{Ade:2015xua}. The various peaks in Fig.~\ref{fig:density} are respectively (from left to right) due to the SM $Z$ and Higgs bosons, $\Delta_R^0$ and $Z_R$.
We find that a TeV scale RHN could accommodate the observed DM relic density, with the annihilation dominated by the heavy scalar and/or $Z_R$ bosons, depending largely on the quartic coupling $\lambda$, the gauge coupling $g_R$ and the $R$-sector VEVs $v_R$, $w_R$.
When the RHN is light, i.e. $M_N \lesssim \mathcal{O} (100 \, {\rm GeV})$, it could easily overclose the Universe even at the SM $h$ ($Z$) resonance, as the coupling of $h$ ($Z$) to DM is heavily suppressed by the mixing angle $\zeta_S$ ($\zeta_Z$). One should note in Fig.~\ref{fig:density} for a fixed $Z_R$ mass, the DM $N$ can not be arbitrarily heavy, as $M_N$ and $M_{Z_R}$ are both proportional to the RH scales and $M_N / M_{Z_R} \sim f' / g_R$. In the very high mass limit, the annihilation cross section scales as $\sim M_N^{-2}$, until it hits the unitarity bound at sufficiently high energy scale, as discussed in Section~\ref{sec:3.3}.

\subsection{Going beyond the unitarity bound} \label{sec:3.3}

For generic range of parameters of the model, as the DM mass is increased beyond  ${\cal O}$(100 TeV) or so, the well-known partial wave unitarity limit kicks in~\cite{kamion}.
To see this explicitly in our model, we write down the thermal averaged annihilation cross section (cf. Appendix~\ref{app:A}) at leading order in $v_{\rm EW}^2 / v_R^2$:
\begin{eqnarray}
\label{eqn:sigmav_pev}
\langle \sigma v \rangle
&=& \frac{3 f^{\prime 2} \lambda^2}{1024 \pi v_R^2}
\left| 1 - \frac{4M_N^2}{4M_N^2-M_{\Delta_R^0}^2 + i M_{\Delta_R^0} \Gamma_{\Delta_R^0}} \right|^2 \langle v^2 \rangle
+ \frac{\tilde{g}^4 M_N^2}{4 \pi M_{Z_R}^4}
\left( 1 - \frac14 \langle v^2 \rangle \right) \nonumber \\
&& + \frac{5 g_R^4 \tan^4\phi}{384 \pi M_N^2}
\left| \frac{4M_N^2}{4M_N^2-M_{Z_R}^2 + i M_{Z_R} \Gamma_{Z_R}} \right|^2 \langle v^2 \rangle \nonumber  \\
&& + \frac{\tilde{g}^4 M_N^2 }{24 \pi M_{Z_R}^4}
\left| 1- \frac{4M_N^2}{4M_N^2-M_{Z_R}^2 + i M_{Z_R} \Gamma_{Z_R}} \right|^2 \langle v^2 \rangle \,,
\end{eqnarray}
where $\tilde g^4 \equiv g_L^2 g_R^2 \xi_Z^2 / (16 \cos^2 \theta_W \cos^2 \phi)$. 
Thus the cross section is suppressed by the right-handed scale $v_R$, i.e. $\langle \sigma v \rangle \propto v_R^{-2}$, except when $2M_N\simeq M_{\Delta_R^0}$ or $M_{Z_R}$, which results in a Breit-Wigner enhancement. Even in this case, we must ensure that the maximum value of the cross section at the resonance obeys the partial wave unitarity limit~\cite{Nussinov:2014qva}. For the resonance $R(=\Delta_R^0, Z_R)$ just above the threshold $2M_N\lesssim M_R$, Eq.~\eqref{eqn:sigmav_pev}  can be approximated by
\begin{eqnarray}
\langle \sigma v \rangle \ \simeq \ 16 \pi \, \frac{\Gamma_{NN}\Gamma_{\rm SM}}{(4M^2_N-M^2_R)^2+\Gamma^2_{R}M^2_R}
\ \sim \  \frac{4\pi} {M_N^2} \left( B_{NN}B_{\rm SM} \right) \,,
\label{eq:uni2}
\end{eqnarray}
where $\Gamma_{NN}$ and $\Gamma_{\rm SM}$ are the partial decay widths of $R$ to the DM pair and SM particles respectively, and $B_{NN}$ and $B_{\rm SM}$ are the corresponding branching ratios. From Eq.~\eqref{eq:uni2}, we find that the annihilation rate decreases with increasing DM mass, which leads to the unitarity bound of $\sim 20$ TeV in our model~\cite{Dev:2016qeb}. Our goal in this subsection is to point out that in our model, there is a way to relax this generic bound without resorting to fine tuning of parameters (as required,~e.g.~in using an $s$-channel resonance below the threshold when $2M_N\gtrsim M_{\Delta_R^0}$~\cite{dkmtz}). The basic idea is that in our model, there are naturally occurring long lived colored fermions (the $SU(2)_R$ quarks) which are heavy e.g. the lightest heavy $SU(2)_R$ sector quark, or the next to lightest RH neutrino, $N_2$. We will show that there is a range of parameters in the model where, these long lived particles (denoted by $X=Q, N_2$) can decay to relativistic lighter SM fermions below the freeze-out temperature of the RHN DM. In that decay process sufficient entropy release can occur leading to dilution of the DM density to the right level.\footnote{Late decays of heavy particles have been used in order to dilute DM abundance in other contexts~\cite{Bezrukov:2009th, miha, Zavala:2014dla, Nussinov:2014qva, Babu:2014uoa}.} 

There is a range of parameters of the model, where the heavy particle $X$ decays at temperature $T_X$ to relativistic  species after $N_1$ relic density is frozen (roughly around $T_F\sim M_{N_1}/20$) i.e. $T_X< T_F$. It will then generate entropy which can dilute the relic density of $N_1$. There is also some dilution each time heavy species annihilate and disappear from the cosmic soup. 
The $X$ decay temperature is given by
\begin{eqnarray}
T_X \ \simeq \ 0. 78 g_*^{-1/4}\sqrt{\Gamma_{X}M_{\rm Pl}} \, .
\end{eqnarray}
We  calculate $\Gamma_{X}$ for each choice of $X=Q, N_2$ and verify that $X$ decays after $T_F$ by adjusting the mixing of heavy quarks to light quarks $\delta_{U, D}$ and the mass $M_X$. The dilution factor is then calculated as follows: equating the energy density before and after the decay we get
\begin{eqnarray}
Y_{X}M_{X} s_{\rm before} \ = \ \frac{3}{4}s_{\rm after}T_X \, ,
\end{eqnarray}
where $Y_X=n_X/s$ is the number density of $X$ normalized to the entropy density.
The dilution factor is given by
\begin{eqnarray}
d \ \equiv \ \frac{s_{\rm after}}{s_{\rm before}} \ = \ \frac{4}{3}~\frac{Y_{X}M_{X}}{0. 78 g_*^{-1/4}\sqrt{\Gamma_{X}M_{\rm Pl}}} \, .
\end{eqnarray}

As an illustration, we show below typical parameter values for the case of dilution by $N_2$.
We choose $M_{N_2}\sim 10$ PeV and heavier quarks coupled to $W_R$ to have masses such that $N_2$ decay to them is kinematically suppressed/forbidden. The decay of $N_2$ then takes place via the mixings $\delta_{U,D}$ of the heavy quarks to the lighter ones. We find that if $\delta_U \delta_D\sim 10^{-3}$ GeV$^2$ we can get $T_{N_2}\sim 10$ GeV (i.e. above the QCD phase transition scale). Taking the annihilation cross section for $N_2N_2\to $ SM fermions
\begin{eqnarray}
\sigma_{N_2N_2} \ \simeq \ \frac{\alpha^2_{Z_R}M^2_{N_2}}{M^4_{Z_R}} \, ,
\end{eqnarray}
where $\alpha_{Z_R}\equiv C_{Z_R N N}^2/4\pi$ (with $C_{Z_RNN}$ defined in Appendix~\ref{app:A}),
we estimate $Y_{N_2}M_{N_2}$  at the time of $N_2$ decay which gets converted to entropy at $T_{N_2}$ leading to a dilution factor $d \simeq 100$ for $\alpha_{Z_R}\simeq 10^{-2}$.

For the heavy quarks, due to QCD couplings being much larger than $Z_R$ coupling to $N_2$, we find that the dilution factor from heavy quark decay is not very strong (roughly of order 2--3). There is also some dilution of DM relic density due to increase in entropy at the QCD phase transition point. Overall, it is possible to get a total dilution factor as big as $d\sim 10^6$, which can relax the unitarity bound for $M_N$ up to a PeV or so~\cite{Dev:2016qeb}.

\section{Direct detection} \label{sec:4}

With the scalars $h$, $\Delta_R^0$ and gauge bosons $Z$, $Z_R$ connecting the RHN DM to the SM sector, the RHN $N$ can scatter off the target nuclei with an observable rate in DM direct detection experiments. As a Majorana DM candidate, the RHN has both spin-independent (SI) and spin-dependent (SD) interactions with nuclei, which are mediated by the scalars and gauge bosons respectively.

\begin{figure}
  \centering
  \includegraphics[width=0.6\textwidth]{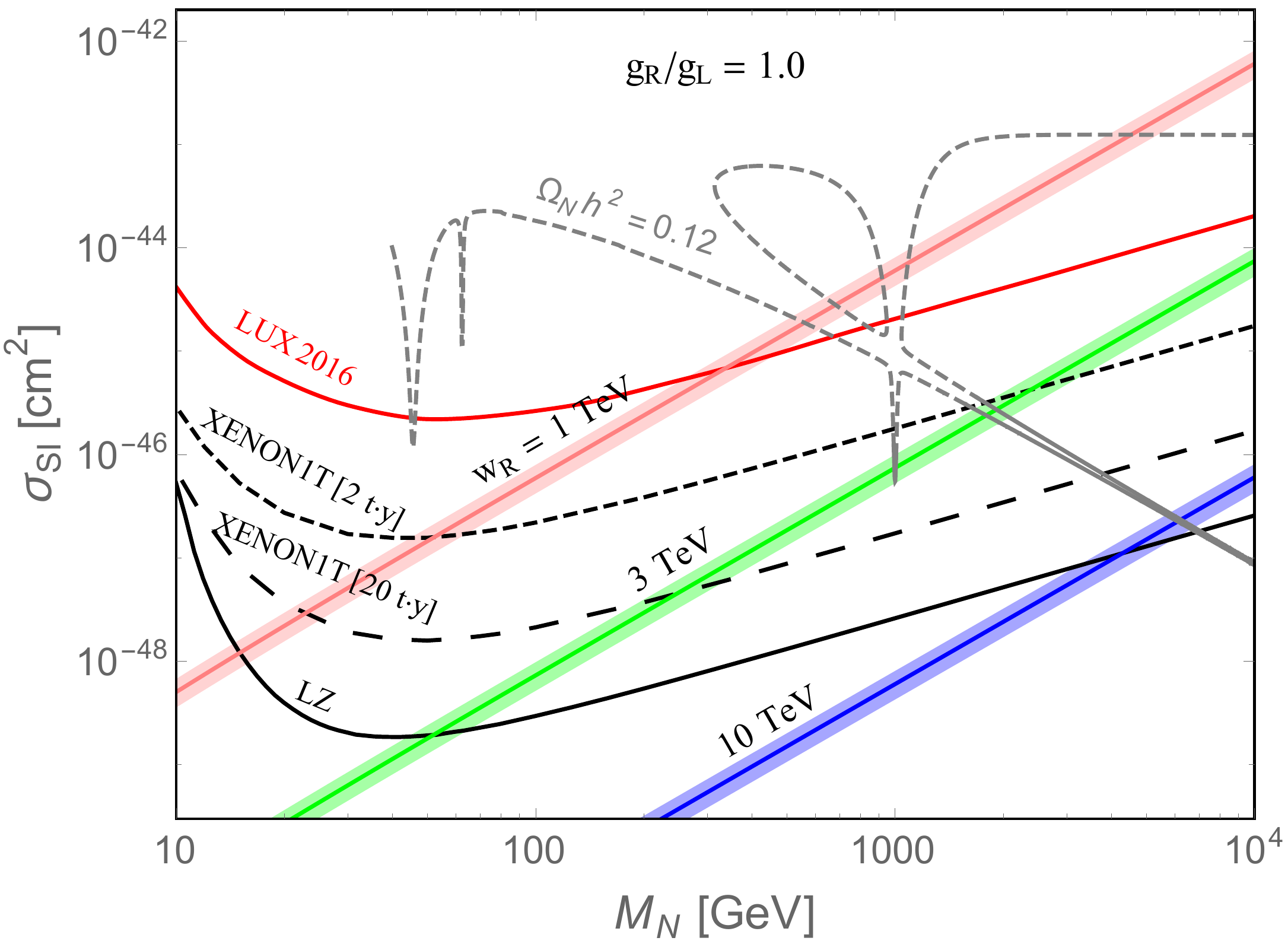}
  \caption{Predictions for the SI DM-nucleon scattering cross sections as functions of DM mass $M_N$ with the gauge coupling $g_R = g_L$ and the RH scale $w_R = 1$, 3 and 10 TeV. The colorful bands are due to the sizable uncertainties of effective DM-nucleon couplings $f_{p,\,n}$ [cf. Eq.~\eqref{eq:fpn}]. We show also the current upper limits from LUX~\cite{Akerib:2016vxi}, as well as the future reaches of XENON1T~\cite{Aprile:2015uzo} and LZ~\cite{Akerib:2015cja}. The gray curves correspond to the parameter space producing the observed relic density $\Omega_N h^2 = 0.12$ (with the dashed part excluded by direct collider searches of $Z_R$ in the dilepton channel, while the solid part is consistent with all constraints).}
  \label{fig:dd1}
\end{figure}

The SI scattering cross section reads, at the zero-momentum transfer limit,
\begin{eqnarray}
\label{eqn:sigmaSI}
\sigma_{\rm SI} \ = \
\frac{\lambda^2 \mu^2 M_N^2}{2\pi m_h^4 w_{R}^4 }
\left[ Z f_p^{} + (A-Z) f_n^{} \right]^2
\end{eqnarray}
where $\mu = M_N M_{\rm nuclei} / (M_N + M_{\rm nuclei})$ is the reduced mass of DM-nucleus system, $f_{p,n}=m_{p,n} ( \frac29+ \frac79  \sum_{q=u,d,s} f^{p,n}_{T_q})$ the effective DM-protron/neutron couplings, with the numerical magnitudes of the parameters $f^{p}_{T_q}$ defined as $m_{p}f^{p}_{T_q}\equiv \langle p|m_q\bar{q}q|p\rangle$ (and similarly for $n$)~\cite{Ellis:2000ds}:\footnote{These values are in good agreement with those extracted from an effective field theory approach~\cite{Alarcon:2011zs, Alarcon:2012nr}.}
\begin{eqnarray}
\begin{array}{lll}
f^p_{T_u} = 0.020\pm 0.004 \,,\; & f^p_{T_d} = 0.026\pm0.005 \,,\; & f^{p}_{T_s} = 0.118\pm0.062 \,, \\
f^n_{T_u} = 0.014\pm0.003 \,, \;\; & f^n_{T_d} = 0.036\pm 0.008 \,,\;\; & f^{n}_{T_s} = 0.118\pm0.062 \,.
\end{array}
\label{eq:fpn}
\end{eqnarray}
In Eq.~\eqref{eqn:sigmaSI}, we include only the SM $h$ mediator, and neglect the heavy $\Delta_R^0$ scalar, which is comparatively suppressed by its large mass $M_{\Delta_R^0}^4 / m_h^4$. A  characteristic feature of our model is that the SI direct detection cross section scales with the DM mass square (as long as $M_N \gg M_{\rm nuclei}$), i.e. $\sigma_{\rm SI} \propto M_N^2$, which originates from the dependence of Yukawa coupling on the DM mass $f' \propto M_N$. The dependence of $\sigma_{\rm SI}$ on the RH scale $w_R$ is from the $h - \Delta_R^0$ mixing angle $\zeta_S \propto w_R^{-1}$ and the Yukawa coupling $f' \propto w_R^{-1}$. This nontrivial dependence is explicitly shown in Fig.~\ref{fig:dd1}, where we set the RH VEVs $v_R = w_R =$ 1, 3 and 10 TeV, the gauge coupling $g_R = g_L$ and the quartic coupling $\lambda = 1$. The colorful bands are due to the sizable uncertainties of effective DM-nucleon couplings $f_{p,\,n}$ [cf. Eq.~\eqref{eq:fpn}]. The SI DM-nucleon scattering cross section is severely constrained, and currently the most stringent limit comes from the LUX experiment~\cite{Akerib:2016vxi}. (with comparable limits from PandaX-II~\cite{Tan:2016zwf} up to 1 TeV DM mass).
The future projected limits from XENON1T (with two exposure values of 2 and 20 ton$\cdot$year)~\cite{Aprile:2015uzo} and LZ~\cite{Akerib:2015cja} are also shown in Fig.~\ref{fig:dd1} for comparison with the model predictions. The gray curves indicate the parameter space with the right relic density $\Omega_N h^2 = 0.12$, whereas the dashed gray part is excluded by direct LHC searches of $Z_R$ in the dilepton channel (cf. Section~\ref{sec:3.1}). Note that for a given DM mass, it is possible to have multiple solutions for the correct relic density depending on the $Z_R$ and $\Delta_R^0$ masses (as in Fig.~\ref{fig:density}, where there are two solutions on either side of each of the $\Delta_R^0$ and $Z_R$ resonances). Although the current direct detection limits are not stringent enough to probe the allowed parameter space of the model, the upcoming ton-scale experiments will have the sensitivity to probe part of this region, with the RH scale $w_R$ up to 10 TeV, far beyond the capability of direct production at LHC.

\begin{figure}
  \centering
  \includegraphics[width=0.6\textwidth]{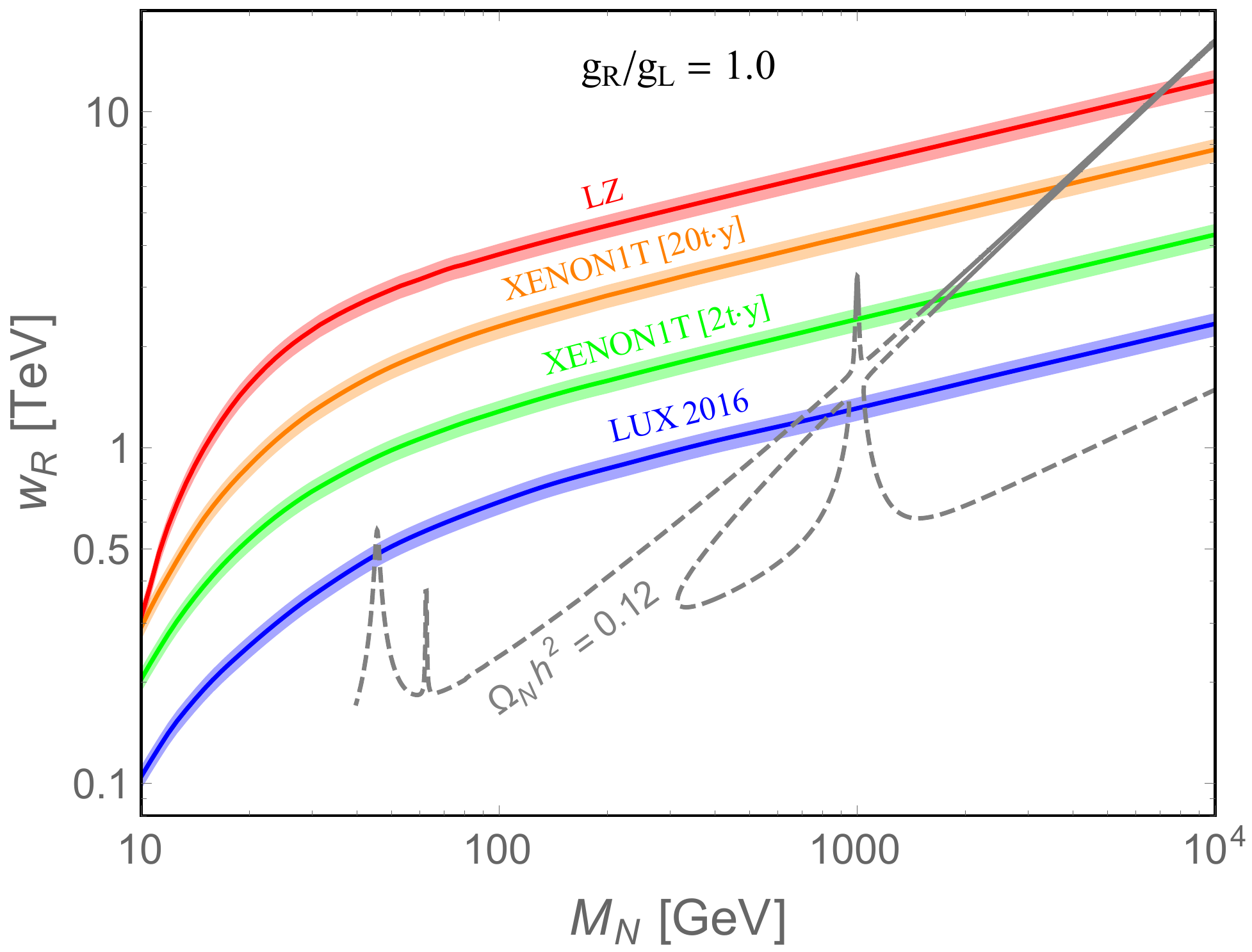}
  \caption{Current and future experimental limits on the SI scattering of the DM $N$ off nucleon, expressed as {\it lower} limits on the rescaled RH scale $w_R$ as a function of the DM mass $M_N$ with gauge coupling $g_R  = g_L$. See text and the caption of Fig.~\ref{fig:dd1} for more details.}
  \label{fig:ddlimit1}
\end{figure}

It is instructive to recast the SI limit shown in Fig.~\ref{fig:dd1} onto constraints on the RH scale $w_R$ in Eq.~(\ref{eqn:sigmaSI}) as a function of the DM mass $M_N$, which is presented in Fig.~\ref{fig:ddlimit1}. The parameter space below the colored blue curve in Fig.~\ref{fig:ddlimit1} is excluded by LUX, which sets a lower limit on the RH scale $w_R$ for any given value of the DM mass $M_N$. When combined with the relic density constraint, this gives a lower limit on the DM mass of order 1 TeV, irrespective of the collider constraint on $Z_R$. Note that for the relic density curves in Fig.~\ref{fig:ddlimit1}, the vertical peaks (from left to right) correspond to the SM $Z$ and Higgs bosons and the $\Delta_R$ resonances, whereas the slanted peak corresponds to the $Z_R$ resonance, whose location depends on the RH scale $w_R$. As expected from Eq.~(\ref{eqn:sigmaSI}), the direct detection constraints become more stringent when the DM becomes heavier.  It is interesting to note that the future direct detection experiments are sensitive to the multi-TeV scale DM in our model, even up to 10 TeV, thus complementing the direct LHC searches. In this plot we set explicitly the quartic coupling $\lambda=1$. For a coupling $\lambda \neq 1$, the constraints in Fig.~\ref{fig:ddlimit1} should be rescaled by a factor of $\sqrt{\lambda}$, as indicated in Eq.~(\ref{eqn:sigmaSI}).

\begin{figure}
  \centering
  \includegraphics[width=0.6\textwidth]{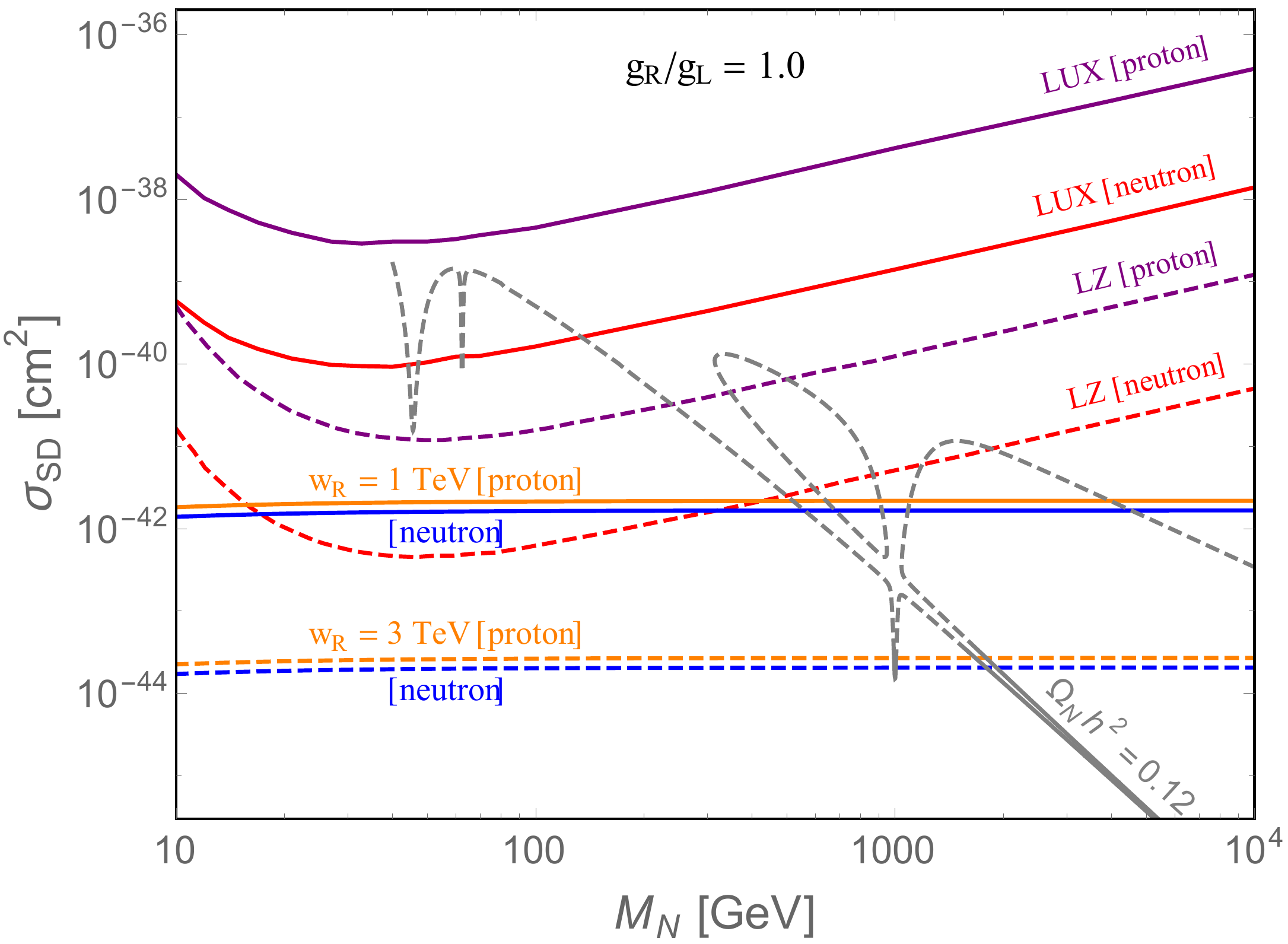}
  \caption{Predictions for the SD scattering cross sections off neutron (blue horizontal line) and proton (orange horizontal line) assuming the gauge coupling $g_R = g_L$ and $w_R = 1$ and 3 TeV. We show also the current limit from LUX~\cite{Akerib:2016lao} and future limit from LZ~\cite{Akerib:2015cja}. See text and the caption of Fig.~\ref{fig:dd1} for more details.}
  \label{fig:sd1}
\end{figure}

As for the SD cross sections, the leading order SD scattering of RHN DM from nuclei is from the axial-vector coupling of SM $Z$ boson and $Z_R$ boson to partons in the detector nuclei. The $Z$ portal is suppressed by the $Z - Z_R$ mixing $M_Z^2/M_{Z_R}^2$ while the $Z_R$ channel by its mass $M_{Z_R}$. It turns out that the two contributions are eventually of the same order, given by
\begin{eqnarray}
\label{eqn:sigmaSD}
\sigma_{\rm SD} \ = \ \frac{ 64 \mu^2 g_R^2 } { \pi m_Z^4 }
\left( \frac{v_{\rm EW}}{v_R} \right)^4
\left[ \sum_{q=u,d,s}   (g_Z^A)_q^{}\lambda_q^{} \right]^2 J_N^{} (J_N^{}+1) \,,
\end{eqnarray}
where $J_N$ is the total angular momentum quantum number of the nucleus, which equals to $1/2$ for free nucleons, and $\lambda_q$ depends on the nucleus and reduces to the light quark contributions $\Delta_q^p \: (\Delta_q^n)$ for scattering off free proton (neutron)~\cite{Airapetian:2006vy}:
\begin{eqnarray}
\Delta^p_u \ = \ \Delta^n_d \ = \ 0.842 \,, \;\;
\Delta^p_d \ = \ \Delta^n_u \ = \ - 0.427 \,, \;\;
\Delta^p_s \ = \ \Delta^n_s \ = \ - 0.085 \,,
\end{eqnarray}
where we neglect the uncertainties on the effective DM-nucleon couplings. The predictions of SD cross sections are presented in Fig.~\ref{fig:sd1}, where we set $g_R = g_L$ and $w_R = 1$, 3 TeV. When the collider constraints on $M_{Z_R}$ are taken into consideration, the $Z - Z_R$ mixing is so  small that even with the future LZ limits it is rather challenging to test the model in terms of SD scattering, unlike the case of SI scattering, where large quartic couplings can enhance the testability of the model.

\begin{figure}
  \centering
  \includegraphics[width=0.49\textwidth]{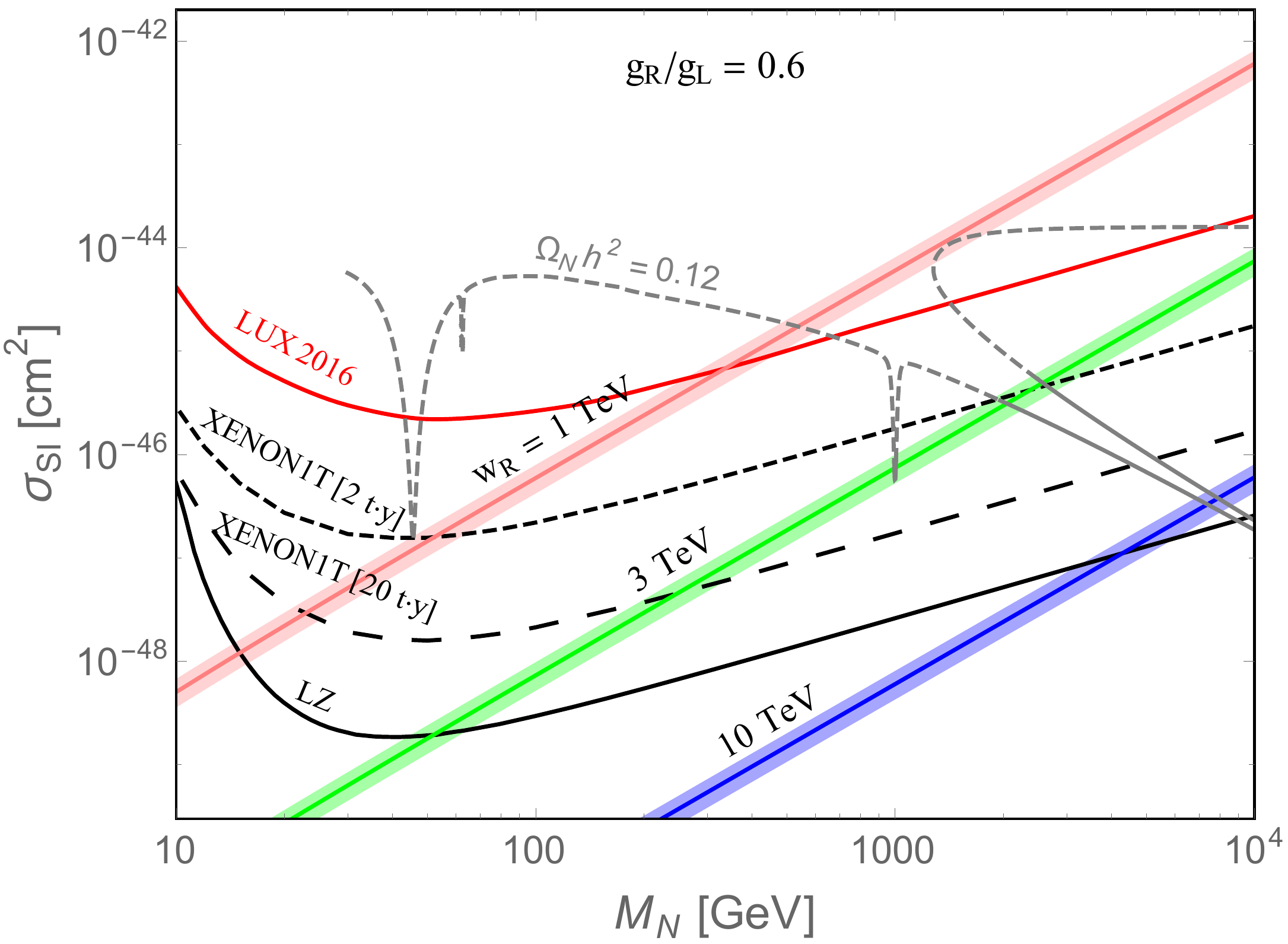} \hspace{-10pt}
  \includegraphics[width=0.49\textwidth]{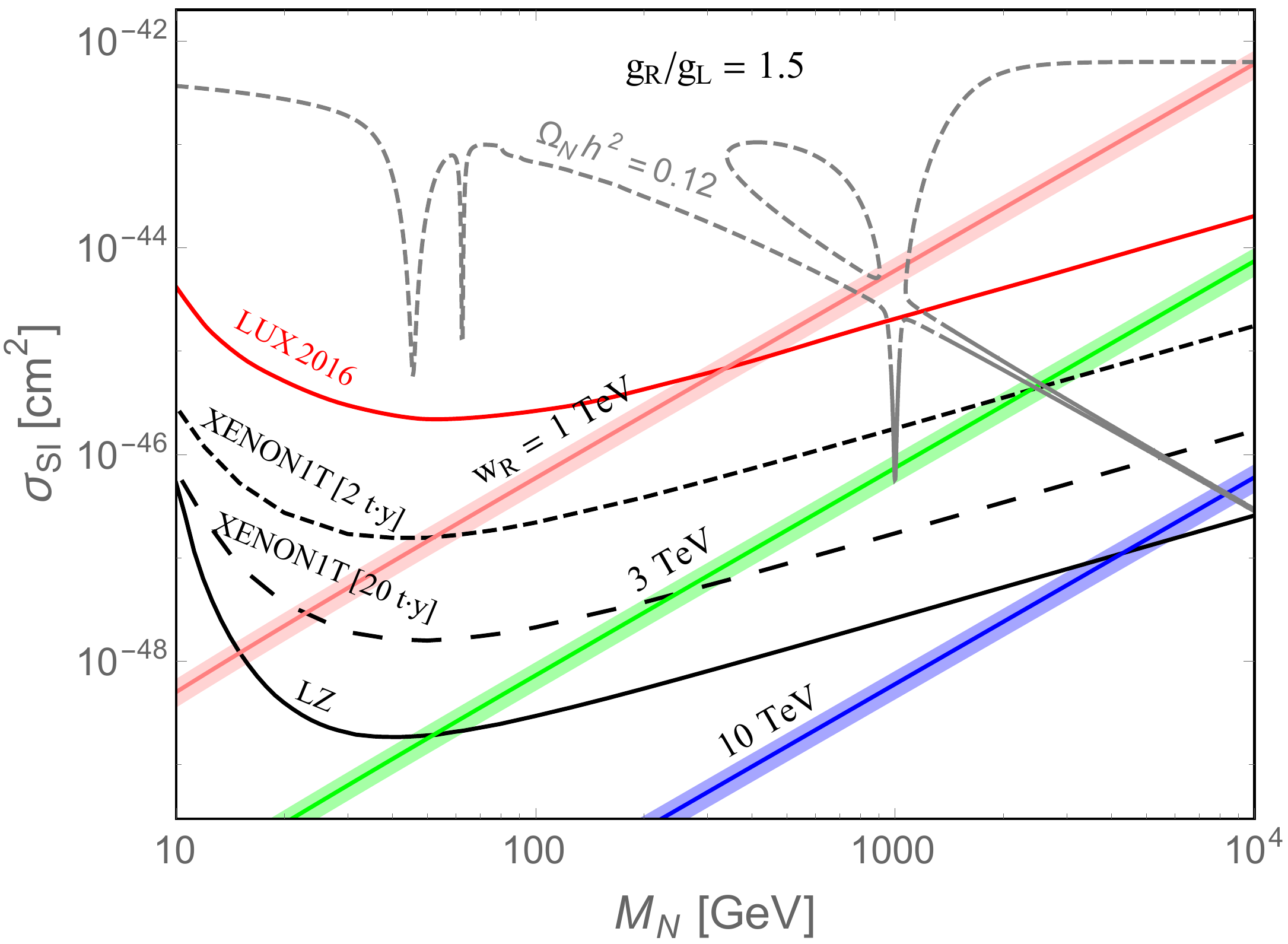}
  \caption{Same as in Fig.~\ref{fig:dd1}, but with $g_R / g_L = 0.6$ (left panel) and 1.5 (right panel).}
  \label{fig:dd2}
\end{figure}

\begin{figure}
  \centering
  \includegraphics[width=0.49\textwidth]{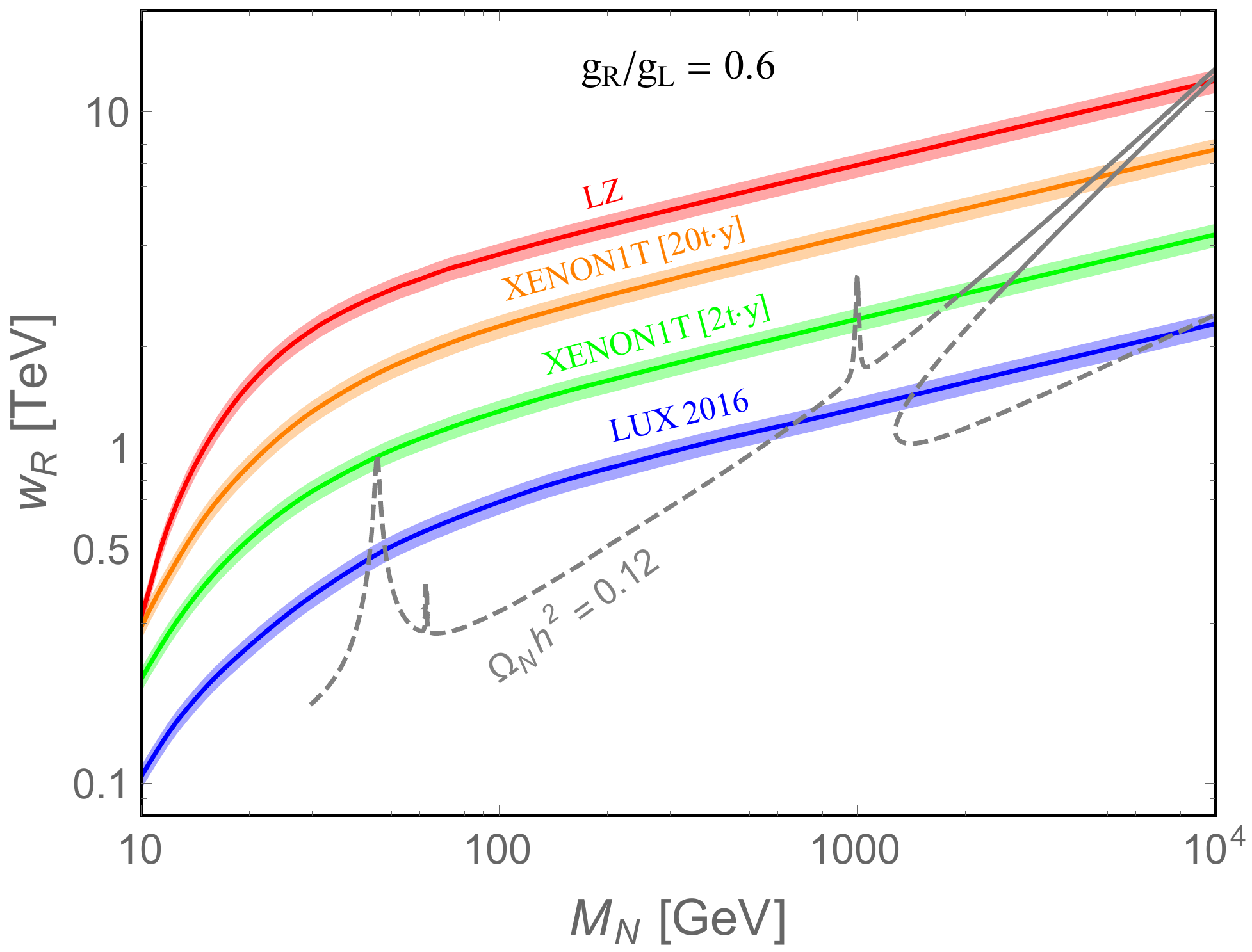} \hspace{-10pt}
  \includegraphics[width=0.49\textwidth]{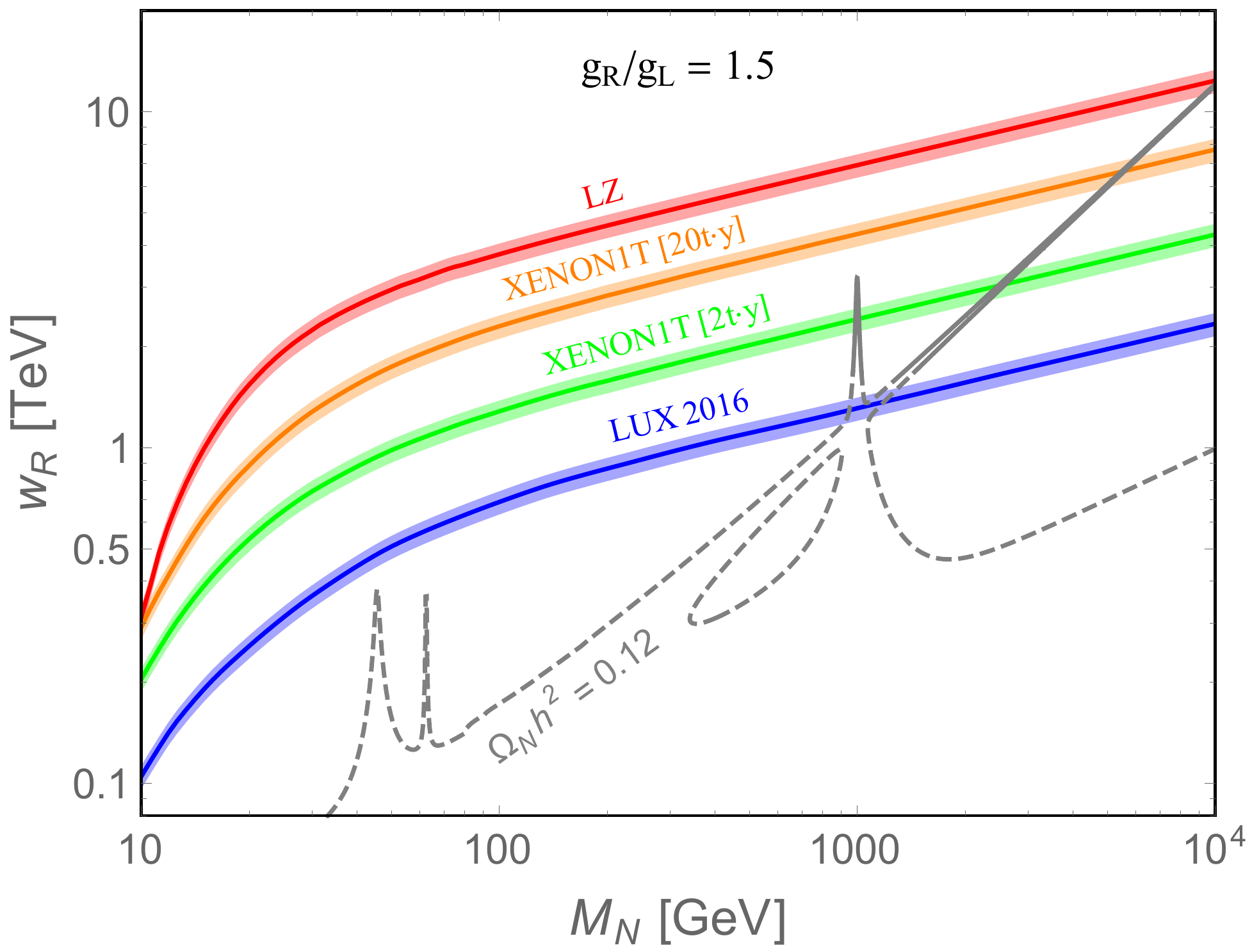}
  \caption{Same as in Fig.~\ref{fig:ddlimit1}, but with $g_R / g_L = 0.6$ (left panel) and 1.5 (right panel).}
  \label{fig:ddlimit2}
\end{figure}

\begin{figure}
  \centering
  \includegraphics[width=0.49\textwidth]{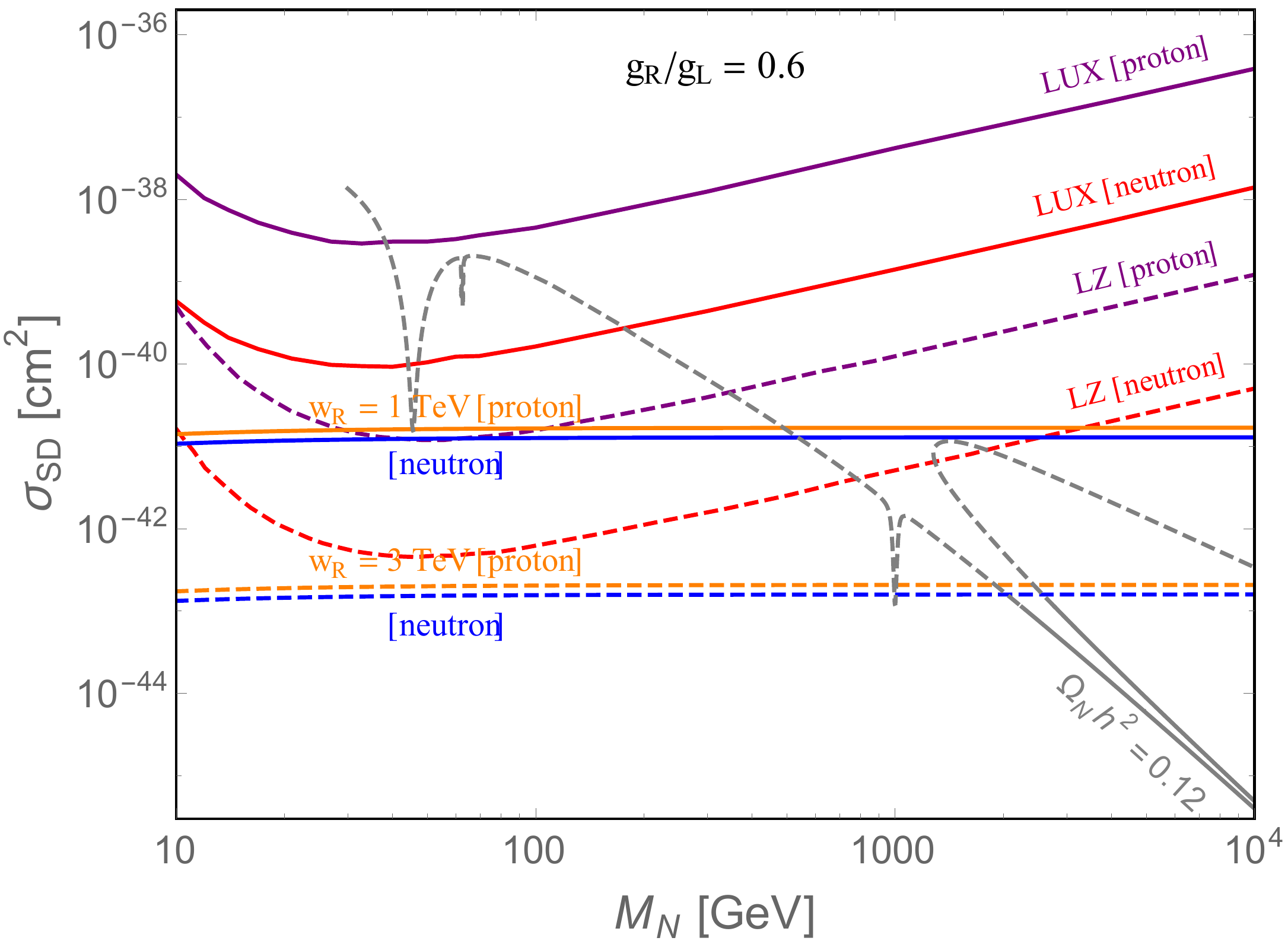} \hspace{-10pt}
  \includegraphics[width=0.49\textwidth]{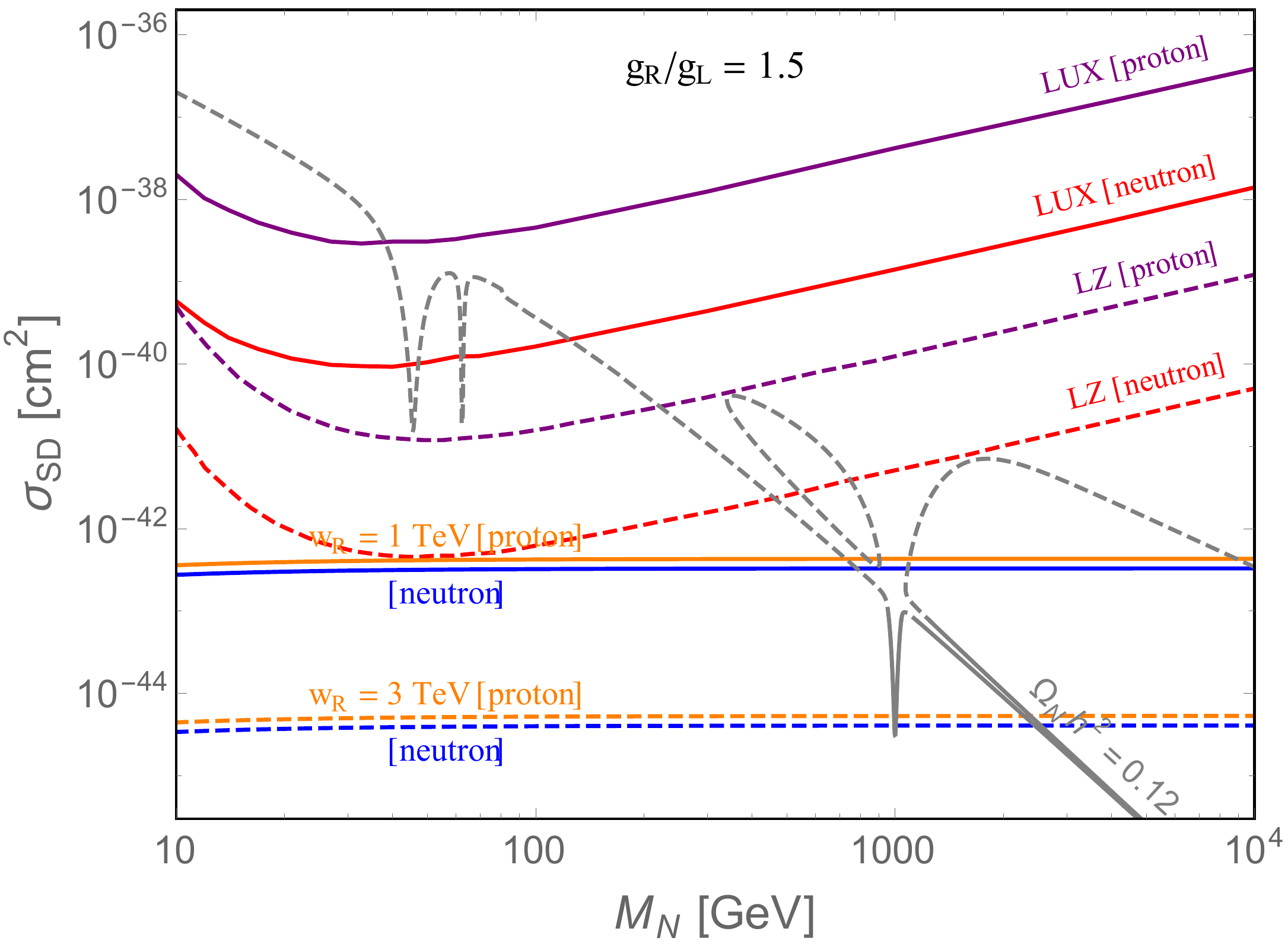}
  \caption{Same as in Fig.~\ref{fig:sd1}, but with $g_R / g_L = 0.6$ (left panel) and 1.5 (right panel).}
  \label{fig:dd3}
\end{figure}

So far in this section we have used the gauge coupling $g_R = g_L$. A smaller or larger $g_R$ is also phenomenologically viable in different classes of LR models. As demonstrative examples, we show the corresponding SI and SD scattering cross section plots with $g_R / g_L = 0.6$ and 1.5 in Figs.~\ref{fig:dd2} and \ref{fig:dd3}, and the lower limits on $w_R$ in Fig.~\ref{fig:ddlimit2}. For the SI scattering, which is from the scalar channel, only the relic density lines $\Omega_N h^2 = 0.12$ are changed and shifted. For a smaller $g_R$, the collider constraints of $M_{Z_R}$ become more stringent (cf. Table~\ref{tab:cs}], and the allowed RH scale $w_R$ is shifted to higher values, as presented in the left panels of Fig.~\ref{fig:dd2} and \ref{fig:ddlimit2}. On the contrary, with a large $g_R$, $w_R$ could be lowered and more parameter space can be probed by the upcoming DM direct detection experiments, as shown by the right panels in Fig.~\ref{fig:dd2} and \ref{fig:ddlimit2}.

The SD DM-nucleon scattering depends directly on the gauge coupling $g_R$ [cf. Eq.~(\ref{eqn:sigmaSD}], and the most significant effect of changing $g_R$ is on the predictions of $\sigma_{\rm SD}$. It is readily understood that for a smaller (larger) $g_R$, the effective coupling of $Z$ and $Z_R$ to the RHN DM $N$ is larger (smaller), and thus the cross section $\sigma_{\rm SD}$ becomes larger (smaller). However, in both the cases shown in Fig.~\ref{fig:dd3}, the cross section $\sigma_{\rm SD}$ for the allowed range of model parameters lies below the future projected limits.



\section{Indirect constraints} \label{sec:5}

\begin{figure}
  \centering
  \includegraphics[width=0.6\textwidth]{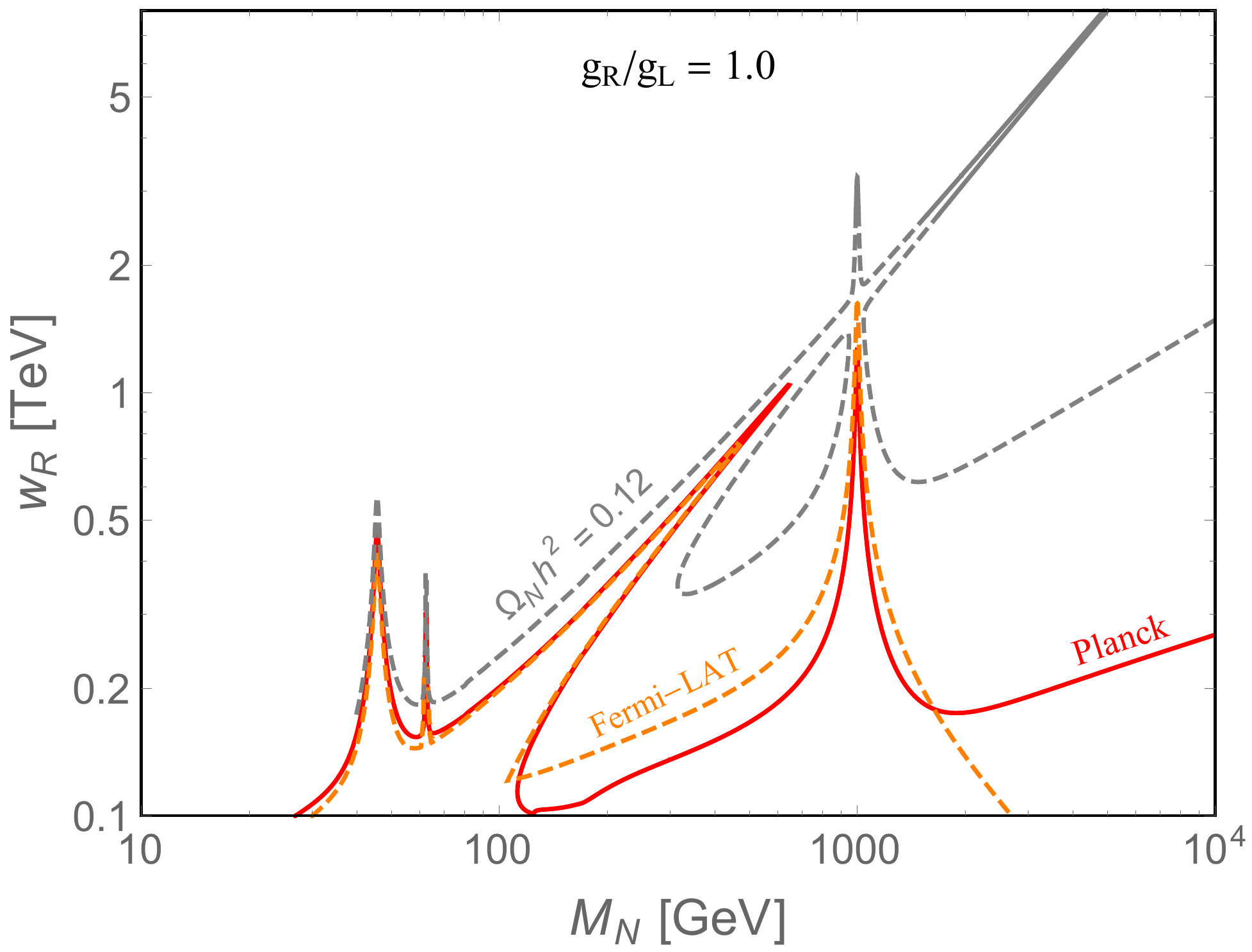} \\
  \caption{Indirect constraints from Fermi-LAT~\cite{Ackermann:2015zua} and Planck~\cite{Ade:2015xua} on the parameter space of $M_N$ and $w_R$ with $g_R = g_L$. The regions below (or inside part of) the colored curves are excluded. The gray curves correspond to the parameter space producing the observed relic density $\Omega_N h^2 = 0.12$ (the dashed parts are excluded by direct collider searches of $Z_R$).}
  \label{fig:idd}
\end{figure}

The RHN DM pairs in our model annihilate in the Universe into electrically charged SM particles, e.g. $\ell^+ \ell^-$, $q \bar q$ and $W^+ W^-$, mediated by the scalar and neutral gauge boson channels [cf. Eq.~\eqref{eq:ann}] which could lead to energetic gamma-rays and thus be constrained by current and future gamma-ray observations. However, due to suppression caused by either small mixing angles $\zeta_{S,\,Z}$ or large masses $M_{Z_R,\: \Delta_R^0}$, the resultant gamma-ray signals are below the current Fermi-LAT sensitivity~\cite{Ackermann:2015zua}, if the RHN DM is to have the observed relic density. Interpreting the Fermi-LAT constraints onto the limits on the model parameter space, we have the orange curve in the $(M_N, w_R)$ plot in Fig.~\ref{fig:idd} where we have set the gauge coupling $g_R = g_L$. In this plot, we show the combined Fermi-LAT constraint taking into account all allowed SM final states relevant to our model. The area below (or inside part of) the curve in excluded. 

\begin{figure}
  \centering
  \includegraphics[width=0.49\textwidth]{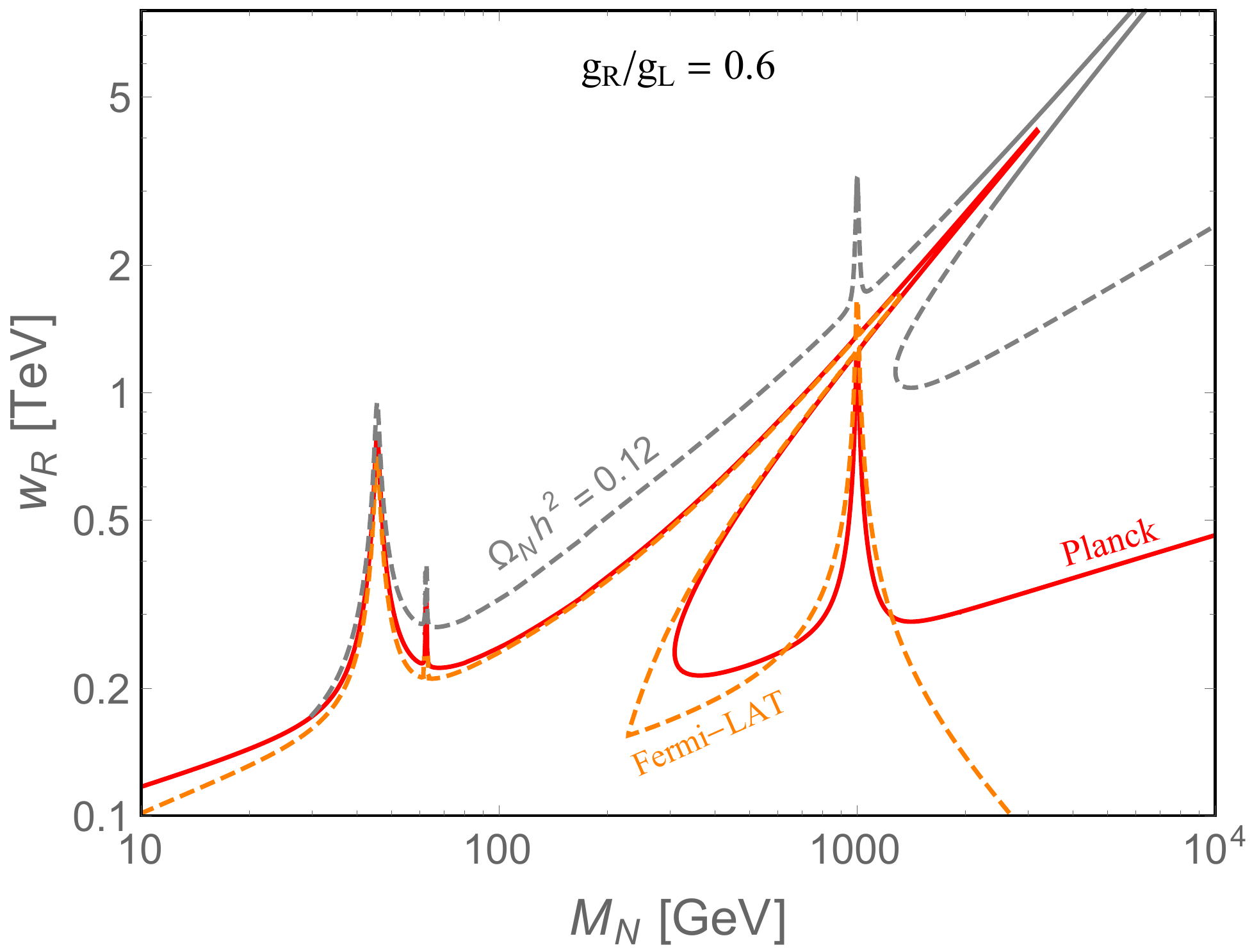}
  \includegraphics[width=0.49\textwidth]{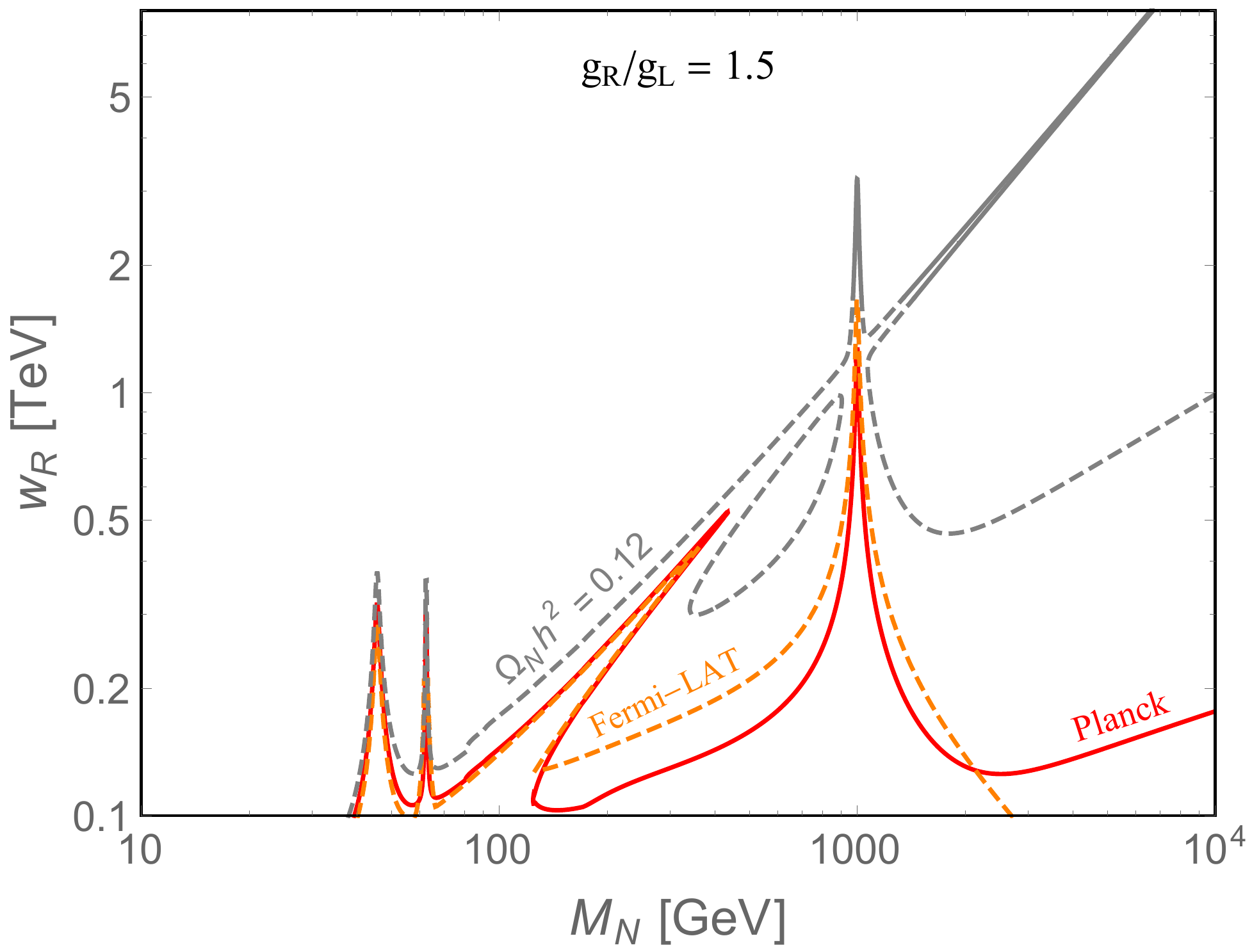}
  \caption{Same as in Fig.~\ref{fig:idd}, but with $g_R / g_L = 0.6$ (left panel) and 1.5 (right panel).}
  \label{fig:idd2}
\end{figure}

On the other hand, DM annihilation injects energy into the thermal bath in the early Universe, thus potentially altering the recombination history and thus changing the
temperature and polarization power spectra of the cosmic microwave background (CMB). The Planck observations~\cite{Ade:2015xua} of CMB anisotropies could therefore constrain the DM annihilation rate, which is complementary to other indirect detections of DM, such as the gamma-rays. The calculation procedure is quite similar to that for the Fermi-LAT limits, with one significant difference here being that we need to rescale the Planck limits given in Ref.~\cite{Ade:2015xua} by the appropriate efficiency factors $f_i(z)$ which depend on the annihilation channels with final states $i$, the redshift $z$ and the DM mass $M_N$. We set the redshift at the epoch of recombination  $z = 1100$ and use the following values of $f_i (z)$ for DM mass of 1 TeV~\cite{Slatyer:2009yq}:
\begin{align}
& f_{u,d,s,c} = 0.349 \,, \quad
f_{b,t} = 0.356 \,, \quad
f_{e} = 0.758 \,, \quad
f_{\mu} = 0.265 \,, \quad
f_{\tau} = 0.244 \,, \nonumber \\
& f_{h} = 0.348 \,, \quad
f_{W} = 0.303 \,, \quad
f_{Z} = 0.287 \,, \quad
f_{hZ} = 0.317 \, .
\end{align}
For the $hZ$ channel we take an average over the $h$ and $Z$ boson factors. For a TeV scale DM, the dependence of $f_i(z)$ on the DM mass is very weak and can be neglected for our illustration purpose. The indirect Planck constraint on the model parameter space is shown by the red curve in Fig.~\ref{fig:idd}, where the area below (or inside part of) the curve is excluded. We find that the Planck limit is comparable to the Fermi-LAT limit, and in some region of the parameter space, even more stringent, in particular for a heavy DM of interest with $M_N \gtrsim 1.5$ TeV.  However, both the Fermi-LAT and Planck limits still cannot constrain any parameter space with the correct relic density. The same conclusion holds  even for a different gauge coupling, as illustrated in Fig.~\ref{fig:idd2} for $g_R/ g_L = 0.6$ and 1.5. Other existing indirect constraints, e.g. from IceCube~\cite{Aartsen:2016exj, Aartsen:2016pfc}, are weaker than the Planck limits, and therefore, not shown here.

\section{Collider phenomenology} \label{sec:6}

As the time-reversed process of annihilation in the Universe, the RHN DM can be pair-produced at high energy colliders, such as the LHC, through the scalar ($h$ and $\Delta_R^0$) and neutral gauge boson ($Z$ and $Z_R$) channels. As the heavy neutrinos are doublets under the gauge symmetry $SU(2)_R$, it can also be produced through the $W_{(R)}$ portal, which is accompanied by a (off-shell) heavy charged lepton ${\cal E}$ in the final state. The heavy lepton ${\cal E}$ decays fast into the DM $N$ plus a charged $W_{(R)}$ boson, which is different from the smoking-gun signal of $W_R$ bosons at colliders in conventional LR models. However, the $W - W_R$ mixing in our model is heavily suppressed by both the loop factor and the soft breaking terms $\delta_{U,\,D}$ (see Section~\ref{sec:2.4}),\footnote{In presence of the mixing term $\delta_\ell \mathcal{E}_L e_R$, the heavy charged lepton $E$ could also transfer into a light SM charge lepton $\ell$ in the final state. However, this term would let the heavy neutrino decay in the early Universe, e.g. $N \to W^\pm \ell^\mp$ through the $W - W_R$ mixing and the charged lepton mixing~\cite{dkmtz}. The RHN DM could also decay into an SM neutrino plus a high energy photon mediated by the charged lepton mixing, although the branching ratio is loop suppressed~\cite{Aisati:2015ova}. For a decaying RHN DM, the lifetime depends on a large variety of soft breaking terms and mixing parameters, and we will not consider the decaying DM scenario in this paper.} thus it is rather challenging to search for the $W_R$ boson in the present model, and we will not consider such $W_{(R)}$ mediated processes in the discussions of DM searches at colliders below.

\subsection{DM searches}



As in case of typical WIMP DM searches at colliders, emission of one or more SM particles from the initial state partons is the smoking-gun collider signal for the RHN DM in our model. Emission of a hard gluon jet from one of the initial partons leads to a monojet plus large missing $E_T$ signal at the LHC, which has been searched for in the $\sqrt s=13$ TeV data~\cite{Aaboud:2016tnv, CMS:2016pod}. There are also DM searches at the LHC in the mono-Higgs~\cite{monoHiggs}, mono-$W/Z$~\cite{monoWZ, CMS:2016hmx, Aaboud:2016qgg} and mono-photon~\cite{CMS:2016fnh} channels. Among these, the monojet channel turns out to be the dominant one for our RHN DM production at hadron colliders.

For pair production of the RHN DM with a high transverse momentum jet, the DM $N$ could interact with the SM fermions through the SM Higgs or $Z$ boson portal. However, both these two channels are from the heavy-light mixing angles $\zeta_S$ or $\zeta_Z$, and thus the dominant channel is through the heavy $Z_R$ boson. Taking into consideration of the current $Z_R$ mass limit discussed in Section~\ref{sec:3.1}, our parton-level simulations reveal that the monojet cross sections in our present model are about three orders of magnitude smaller than the current LHC constraints~\cite{Aaboud:2016tnv}. Thus it is almost hopeless to see the RHN DM directly at the LHC in the monojet channel. Even at future 100 TeV collider~\cite{Golling:2016gvc}, due to the rather low signal to background ratio, it is rather challenging to directly observe the RHN DM with the designed luminosity. The future lepton colliders, e.g. ILC, FCC-ee or CEPC, although much cleaner, have too low colliding energy to set any limits on the RHN DM in our model. So we have to look for alternative collider signatures of our model, as discussed below.

\subsection{Searches for long lived particles at the LHC}
As is clear from the discussion in Section~\ref{sec:2.5}, there is a viable range of the heavy-light quark mixing parameters $\delta_{U,D}$ in our model for which the heavy quarks and resulting heavy baryons are long lived. 
Just to give a feeling for the numbers, we recall the expression for lifetime of a generic heavy quark decay $Q\to qZ$ in terms of the mixing $\delta_{Q}$ as given in Eq.~\eqref{MQq}.
For $M_Q\sim $ 1 TeV, $v_R\sim$ 10 TeV and $\delta_Q\sim 10^{-6}$ GeV, this roughly gives a lifetime $\tau_Q\sim 10^{-7}$ sec or the decay length $L_0\equiv c\tau\sim 30$ m. For higher $\delta_Q$, this distance goes down like $\left({\delta_Q}/{10^{-6} \, {\rm GeV}}\right)^{-2}$ and could give rise to displaced vertices at the LHC. We expect the number of such events to be $\sim N_Q(1-e^{-L/L_0})\epsilon/b$ where $b=|\vec{p}|/M_Q$ is the boost factor for the produced heavy quark, $N_Q$ is the number of heavy quarks produced, $L$ is the distance of the detector from the production point and $\epsilon$ is the detector efficiency. The current searches for displaced vertices~\cite{Rhadron} are roughly sensitive to $\delta_Q\sim 10^{-5}$ GeV.

\begin{figure}
  \centering
  \includegraphics[width=0.49\textwidth]{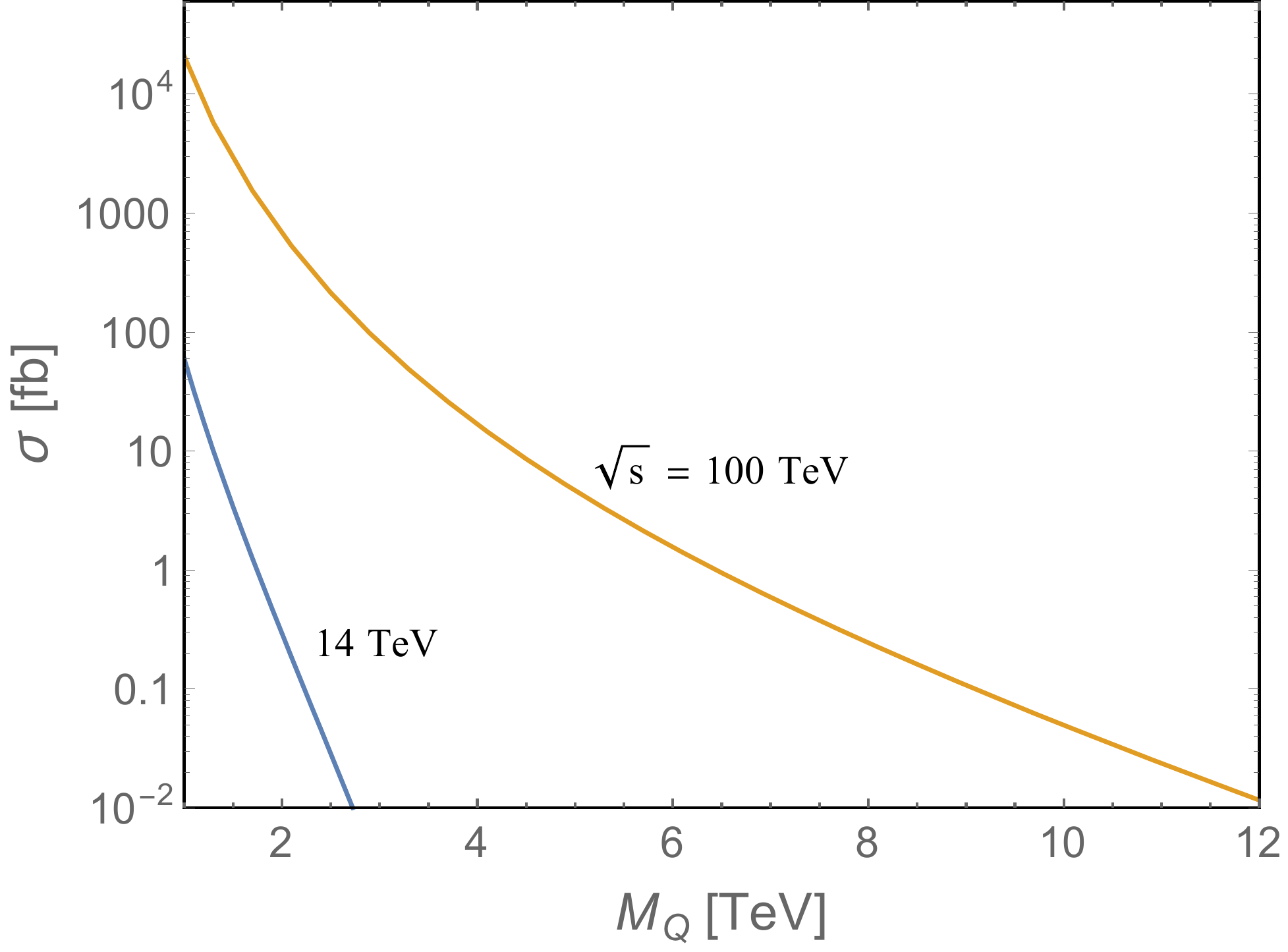}
  \caption{Pair production cross sections of the heavy quarks $Q$ at hadron colliders with $\sqrt{s} = 14$ and $100$ TeV~\cite{Arkani-Hamed:2015vfh}.}
  \label{fig:heavyquark}
\end{figure}

The dominant production for the heavy quarks in our model at hadron colliders is through the processes $gg ,\, q\bar{q} \to Q \bar{Q}$, with the cross sections with $\sqrt{s} = 14$ and $100$ TeV presented in Fig.~\ref{fig:heavyquark}~\cite{Arkani-Hamed:2015vfh}. Note that the single production of heavy quarks in our model is suppressed by the mixing parameter $\delta_Q$. For a 1 TeV heavy quark, we expect a signal number of $N_Q=3.6 \times 10^5$ at 14 TeV with an integrated luminosity of 3000 fb$^{-1}$, and a much larger number of $N_Q=1.2 \times 10^{9}$ at 100 TeV with a luminosity of 30 ab$^{-1}$.

For neutral long-lived particles such as $N_2$ in our model, one may consider a different set-up as has been recently proposed~\cite{david}. 
More details of this will be given elsewhere.

\section{Discussion and Conclusion} \label{sec:7}
Before concluding, we briefly comment on some other phenomenological implication of the model:
\begin{itemize}
\item Even though there are RH neutrinos with coupling to RH gauge boson $W_R$, there are no new dominant contributions to neutrinoless double beta decay and the only observable channels are due to the canonical light neutrino exchange.

\item Since the neutrino mass matrix fixes the Yukawa couplings of the left-handed triplet Higgs field $\Delta_L$ in our model with type-II seesaw, there will be contributions to lepton flavor changing processes such as $\mu\to e\gamma$, $\mu\to 3 e$ etc coming from the $f$-couplings. The severe experimental constraints on these processes impose  constraints on the mass of the members of this triplet to be correlated to the value of $v_L\equiv \langle \Delta^0_L \rangle$. For example, for $f\sim 10^{-2}$, $M_\Delta \gtrsim $ 3 TeV.

\item If needed, our model can naturally accommodate very long lived DM by soft breaking of the $Z_{2,\ell}$ symmetry stabilizing the DM. Depending on the DM mass, this could have additional phenomenological implications, e.g. at IceCube for a PeV-scale decaying DM.



\end{itemize}

In summary, we have discussed the phenomenology and cosmology of a TeV scale DM in the context of an $SU(2)_L\times SU(2)_R\times U(1)_{Y_L}\times U(1)_{Y_R}$ model where there is an automatic $Z_{2,\ell}$ symmetry that guarantees the stability of the DM. As a result, it provides an interesting possibility where the lightest right-handed neutrino $N$ plays the role of cold DM. The DM relic density and direct detection constraints imply a lower limit of order 1 TeV for the RHN DM. The model also predicts the existence of long-lived heavy quarks which can be searched for at the LHC.

\section*{Acknowledgment}
The work of B.D. is supported by the DFG grant RO 2516/5-1. B.D. also acknowledges the local hospitality provided by the CERN Theory Group during the last phase of this work. The work of R.N.M. is supported by the US National Science Foundation Grant No. PHY-1315155. Y.Z. would like to thank the IISN and Belgian Science Policy (IAP VII/37) and the National Natural Science Foundation of China (NSFC) under Grant No. 11375277 for support.  Y.Z. is also grateful to Hong-Hao Zhang for his gracious hospitality during the visit at Sun Yat-Sen University where part of the work was done.

\appendix
\section{Annihilation cross sections} \label{app:A}

The thermally averaged DM annihilation cross section times velocity can be written as
\begin{eqnarray}
\langle \sigma v \rangle = a + b \langle v^2 \rangle + \mathcal{O} (v^4) \,,
\end{eqnarray}
where the $a$ and $b$ coefficients in various channels for our model with DM mass $M_N$ are given below:
\begin{eqnarray}
\label{eqn:a1}
a^{(V,\,A)}_{ff} & = & \frac{N_C^f \beta_f m_f^2 }{2\pi}
 \sum_{i,\,j} M_{V_i}^{-2} M_{V_j}^{-2} \left( C_{V_i NN} C^{(A)}_{V_i ff} \right) \left( C_{V_j NN} C^{(A)}_{V_j ff} \right) \Theta (M_N - m_f) \,, \\
a^{(V)}_{VS} & = & \frac{\beta_{VS}^3 M_N^2 }{16\pi m_V^2}
 \sum_{i,\,j} M_{V_i}^{-2} M_{V_j}^{-2} \left( C_{V_i NN} C_{V_i VS} \right) \left( C_{V_j NN} C_{V_j VS} \right) \Theta (2M_N - m_V - m_S) \,, \\
b^{(S)}_{SS} & = & \frac{\delta_S \beta_S}{128\pi}
\sum_{i,\,j} \left( C_{S_i NN} C_{S_i SS} P_{S_i} \right)
\left( C_{S_j NN} C_{S_j SS} P_{S_j}^\ast \right) \Theta (M_N - m_S) \,, \\
b^{(S)}_{ff} & = & \frac{N_C^f \beta_f^3 M_N^2}{8\pi}
\sum_{i,\,j} \left( C_{S_i NN} C_{S_i ff} P_{S_i} \right)
\left( C_{S_j NN} C_{S_j ff} P_{S_j}^\ast \right) \Theta (M_N - m_f) \,, \\
\label{eqn:b3}
b^{(V,\,V)}_{ff} & = & \frac{N_C^f \beta_f M_N^2 }{6\pi}
\sum_{i,\,j} F_{ff}^{(V,\,V)} \left( C_{V_i NN} C^{(V)}_{V_i ff} P_{V_i} \right)
 \left( C_{V_j NN} C^{(V)}_{V_j ff} P_{V_j}^\ast \right) \Theta (M_N - m_f) \,, \\
\label{eqn:b4}
b^{(V,\,A)}_{ff} & = & \frac{N_C^f \beta_f^{-1} M_N^2}{6\pi}
\sum_{i,\,j} F_{ff}^{(V,\,A)}
\left( C_{V_i NN} C^{(A)}_{V_i ff} P_{V_i} \right)
\left( C_{V_j NN} C^{(A)}_{V_j ff} P_{V_j}^\ast \right) \Theta (M_N - m_f) \,, \\
b^{(S)}_{VV} & = & \frac{\delta_V \beta_V}{128\pi}
\sum_{i,\,j} F_{VV}^{(S)}
\left( C_{S_i NN} C_{S_i VV} P_{S_i} \right)
\left( C_{S_j NN} C_{S_j VV} P_{S_j}^\ast \right) \Theta (M_N - m_V) \,, \\
b^{(V)}_{VV} & = & \frac{ \beta_V M_N^2}{24\pi}
\sum_{i,\,j} F_{VV}^{(V)}
\left( C_{V_i NN} C_{V_i VV} P_{V_i} \right)
\left( C_{V_j NN} C_{V_j VV} P_{V_j}^\ast \right) \Theta (M_N - m_V) \,, \\
b^{(V)}_{VS} & = & \frac{\beta_{VS} M_N^6 }{4\pi m_V^2}
\sum_{i,\,j}  F_{VS}^{(V)} M_{V_i}^{-2} M_{V_j}^{-2}
\left( C_{V_i NN} C_{V_i VS} P_{V_i} \right)
\left( C_{V_j NN} C_{V_j VS} P_{V_j}^\ast \right) \nonumber \\
&& \quad \times \Theta (2M_N - m_V - m_S) \,.
\end{eqnarray}
The subscripts $S$, $f$ and $V$ stand for the SM scalar, fermion and vector final states (all the SM fermion states are assumed to be summed over with the appropriate color factor $N_C$), and the superscripts $S$ and $V$ for the scalar and vector mediators, with the second superscripts $V$ and $A$ in Eqs.~(\ref{eqn:a1}), (\ref{eqn:b3}) and (\ref{eqn:b4}) denoting the ``vector'' and ``axial-vector'' parts of the gauge couplings to SM fermions. Those channels missing for the coefficient $a$ are all vanishing in our model. $\delta_{S,\,V}$ is a symmetry factor, with the value of 1 for identical final states and $2$ for different states. $C_{S_i XX}$ and $C_{V_i XY}$ are the Yukawa and gauge couplings of the mediators to the DM particle $N$ and the SM final states.
In the present model the couplings of SM $h$ and $Z$ bosons to all the SM final state are the same as in the SM at the leading order, while the Yukawa couplings $C_{\Delta_R^0 NN} = \frac12 f'$ and axial-vector current coupling $C_{Z_R NN} = g_R / 2\cos\phi$. The $Z_R$ boson couples to the SM fermions via the couplings $C_{Z_R ff} = -\frac14 g_R \sin\phi \tan\phi (Y_{\rm SM}^{L,f} + Y_{\rm SM}^{R,f})$ with $Y_{\rm SM}^{L,\,R}$ the SM hypercharge for the left- and right-handed fermions.
The couplings of heavy bosons $\Delta_R^0$ to the SM particles and the $h$ and $Z$ couplings to the DM $N$ are essentially from the scalar and vector mixings and thus rescaled respectively by the small mixing angles of $\zeta_S$ and $\zeta_Z$ in Eqs.~(\ref{eqn:mixs}) and (\ref{eqn:mixz}). $P_{X} = (4M_N^2 - M_X^2 + i M_X \Gamma_X)^{-1}$ is the standard propagator for the mediator $X$. The velocities of final states are, at the leading order,
\begin{eqnarray}
\beta_x &\equiv& \left[ 1 - \frac{m_x^2}{M_N^2} \right]^{1/2} \,, \\
\beta_{xy} &\equiv& \left[ 1 - \frac{m_x^2 + m_y^2}{2M_N^2} + \frac{\left( m_x^2 - m_y^2 \right)^2}{16M_N^4} \right]^{1/2}
\end{eqnarray}
and the dimensionless functions are defined as
\begin{eqnarray}
F_{ff}^{(V,\,V)} (M_N, m_f) &\equiv&
1 + \frac{m_f^2}{2 M_N^2}  \,,  \\
F_{ff}^{(V,\,A)} (M_N, M_{V_1}, M_{V_2}, m_f) &\equiv&
1 -\frac{7 m_f^2}{2 M_N^2} + \frac{23 m_f^4}{8 M_N^4}
 + \frac{6 m_f^2}{M_{V_1}^2} + \frac{6 m_f^2}{M_{V_2}^2} +\frac{30 m_f^4}{M_{V_1}^2 M_{V_2}^2} \nonumber \\
&& -\frac{15 m_f^4}{2M_N^2 M_{V_1}^2} -\frac{15 m_f^4}{2M_N^2 M_{V_2}^2}
-\frac{24 M_N^2 m_f^2}{M_{V_1}^2 M_{V_2}^2} \,, \\
F_{VV}^{(S)} (M_N, m_V) &\equiv& 3 - \frac{4M_N^2}{m_V^2} + \frac{4M_N^4}{m_V^4} \,, \\
F_{VV}^{(V)} (M_N, m_V) &\equiv& \left( 1-\frac{m_V^2}{M_N^2} \right)
\left( 3 + \frac{20M_N^2}{m_V^2} + \frac{4M_N^4}{m_V^4} \right) \,, \\
F_{VS}^{(V)} (M_N, M_{V_1}, M_{V_2}, m_V, m_S) &\equiv&
- \left( 1 - \frac{M_{V_1}^2 + M_{V_2}^2}{4M_N^2} + \frac{M_{V_1}^2 M_{V_2}^2}{48 M_N^4} \right) \nonumber \\
&& \times \left[ 1 + \frac{5 (m_S^2+m_V^2)}{4 M_N^2} - \frac{(m_S^2-m_V^2)^2}{4 M_N^4} \right]
\nonumber \\
&& + \frac{M_{V_1}^2 M_{V_2}^2}{32M_N^4} \left[ \frac{m_S^2 + 5m_V^2}{M_N^2} - \frac{(m_S^2-m_V^2)^2}{4M_N^4} \right] \,.
\end{eqnarray}

\end{document}